\documentclass[12pt]{article}
\usepackage{amsmath,amssymb,bbold,epsfig}
\usepackage{amsfonts}
\usepackage{slashed}
\usepackage[normalem]{ulem}
\usepackage{graphicx}
\usepackage[sort&compress,comma,square]{natbib}
\usepackage{bbm}
\usepackage{ctable,longtable}
\usepackage[hypertex]{hyperref}
\newcommand{\appsection}{\addtocounter{section}{1}\setcounter{equation}{0}
                         \renewcommand{\thesection}{\Alph{section}}
}
\renewcommand{\theequation}{\arabic{equation}}
\newcommand{\be}{\begin{equation}}
\newcommand{\ee}{\end{equation}}
\newcommand{\bea}{\begin{eqnarray}}
\newcommand{\eea}{\end{eqnarray}}
\newcommand{\ket}[1]{\ensuremath{| #1 \rangle}}   % Ket vector
\newcommand{\sprod}[2]{\ensuremath{\left\langle #1 |%
                     #2 \right\rangle}}  % QM scalar product
\renewcommand{\vec}[1]{{\mathbf{#1}}}
\newcommand{\MB}{M\"{o}ssbauer}

\begin{document}

\title{\vspace{-2cm}
\hfill {\small FERMILAB-PUB-10-014-T}\\
\vglue 0.8cm
%\hfill {\small arXiv: 1001.YYYY [hep-ph]}
\vskip 0.5cm
\Large \bf
Neutrino oscillations: Quantum mechanics vs. quantum field theory}
\author{
{Evgeny~Kh.~Akhmedov$^{a,b}$\thanks{email: \tt 
akhmedov@mpi-hd.mpg.de}~~\,and Joachim~Kopp $^{c,a}$\thanks{email:
\tt jkopp@fnal.gov}
\vspace*{1.5mm}
} \\
%%%%%%%%%%%%%%%%%%%%%%%%%%%%%%%%%%%%%%%%%%%%%%%%%%%%%%%%%%%%%%%%%
{\normalsize\em $^a$Max-Planck-Institut f\"ur Kernphysik,
Postfach 103980} \\ {\normalsize\em D--69029 Heidelberg, Germany
\vspace*{0.15cm}}
\\
{\normalsize\em $^{b}$National Research Centre Kurchatov
\vspace*{-0.1cm}Institute}\\{\normalsize\em Moscow, Russia 
\vspace*{0.15cm}}
\\
{\normalsize\em $^{c}$Theoretical Physics Department, Fermilab,
                      Batavia, IL 60510, USA}
\vspace*{0.15cm}}
\date{January 28, 2010}  %FIXME
\maketitle
\thispagestyle{empty}
\vspace{-0.8cm}
\begin{abstract}
\noindent
A consistent description of neutrino oscillations requires either the
quantum-mechanical (QM) wave packet approach or a quantum field theoretic (QFT)
treatment. We compare these two approaches to neutrino oscillations and discuss
the correspondence between them. In particular, we derive expressions for the
QM neutrino wave packets from QFT and relate the free parameters of the QM
framework, in particular the effective momentum uncertainty of the neutrino
state, to the more fundamental parameters of the QFT approach. We include in
our discussion the possibilities that some of the neutrino's interaction
partners are not detected, that the neutrino is produced in the decay of an
unstable parent particle, and that the overlap of the wave packets of the
particles involved in the neutrino production (or detection) process is not
maximal.  Finally, we demonstrate how the properly normalized oscillation
probabilities can be obtained in the QFT framework without an ad hoc
normalization procedure employed in the QM approach.   

\end{abstract}

\vspace{1.cm}
%%\centerline{Pacs numbers: .....}
\vspace{.3cm}
%\centerline{Keywords: ...}
%\vspace{.3cm}

\newpage

%=======================================================================
\section{\label{sec:intro}Introduction}
%=======================================================================

It is well known by now that neutrino oscillations can be consistently 
described either in the quantum-mechanical (QM) wave packet approach, or 
within a quantum field theoretic (QFT) framework.%
\footnote{Although in a number of sources a plane wave approach to 
neutrino oscillations is employed, it is actually marred by inconsistencies  
and, if applied correctly, does not lead to neutrino oscillations at all 
\cite{Rich:1993wu,Beuthe1}. } 
In the QM method \cite{Nussinov:1976uw,Kayser:1981ye,Giunti:1991ca,
Rich:1993wu,Kiers,Dolgov:1997xr,Giunti:1997wq,Cardall:1999ze,Dolgov:1999sp,
Dolgov:2002wy,Giunti:2002xg,FarSm,Visinelli:2008ds,AS,nuwp2}, neutrinos 
produced in weak-interaction processes are described by propagating wave 
packets, the spatial length of which is related to the momentum uncertainty 
at neutrino production, and the detected states are also described by wave 
packets, centered at the detection point. The transition (oscillation) 
amplitude is then obtained by projecting the evolved emitted neutrino state 
onto the detection state. In the QFT treatment  
\cite{Kobzarev:1980nk,Kobzarev:1981ra,Giunti:1993se,Rich:1993wu,Grimus:1996av,
Grimus:1998uh,Ioannisian:1998ch,Cardall:1999ze,Dolgov:1999sp,Beuthe2,
Dolgov:2002wy,Dolgov:2004ut,Dolgov:2005nb,AKL1}, one considers neutrino  
production, propagation and detection as a single process, described 
by a tree-level Feynman diagram with the neutrino in the intermediate state 
(see fig.~\ref{fig:feyn}). Neutrinos are represented in this framework by 
propagators rather than by wave functions. 
Both approaches lead to the standard formula for the probability of neutrino 
oscillations in vacuum in the case when the decoherence effects related to 
propagation of neutrinos as well as to their production and detection can be 
neglected. They differ, however, in the way they account for 
possible decoherence effects, with the QFT approach leading 
to a more consistent and accurate description. 
The QM method treats neutrino energy and momentum uncertainties responsible 
for these effects in a simplified way; in addition, it involves an ad hoc 
normalization procedure for the transition amplitude that is not properly 
justified. 

The goal of the present paper is to compare the two approaches and 
establish a relationship between them, as well as to clarify some of the 
procedures that are employed in the QM method from the more general and 
consistent QFT standpoint. Some work in that direction has been done 
before. In \cite{Giunti:2002xg} neutrino wave packets were derived 
starting from the QFT formalism (see also a discussion in 
\cite{Beuthe1}). In~\cite{Kopp:2009fa}, a comparison of the QM and QFT 
approaches was presented for the special case of \MB\ neutrinos, i.e.\ 
neutrinos produced and detected recoillessly in hypothetical \MB-type 
experiments (see, e.g.,~\cite{Raghavan:2005gn,AKL1,Potzel:2008xk} and 
references therein).
The new results obtained in the present work include a more advanced and 
general study of the QFT-based derivation of the neutrino wave packets 
(including the possibility that some of the external particles are not 
detected), matching of the QFT and QM expressions for the neutrino wave 
packets, study of the general properties of the wave packets describing 
the neutrino states (including their energy uncertainties in the case 
when neutrinos are produced in decays of unstable particles) and 
clarification of the issue of 
normalization of the oscillation probabilities in QM and QFT.

The paper is organized as follows. To make the presentation self-contained, in 
Secs.~\ref{sec:QM} and~\ref{sec:QFT} we review, respectively, the QM wave 
packet formalism and the QFT approach to neutrino oscillations. Sections 
\ref{sec:QMQFT} -- \ref{sec:addit} contain our main results. In 
Sec.~\ref{sec:QMQFT} we discuss how the neutrino wave packets, which are 
a necessary ingredient of the QM approach, can be derived starting from the 
QFT formalism. Next, we consider some general properties of the neutrino 
wave packets and discuss the conditions under which they can be approximated 
by Gaussian wave packets. Using the case of Gaussian wave packets as an 
example, we then discuss how the QFT-derived wave packets can be represented 
in the form usually adopted in the QM treatment. We also find expressions for 
the effective parameters describing the QM wave packets in terms of the more 
fundamental input parameters of the QFT framework. Next, we discuss the 
neutrino energy uncertainty in the case when neutrinos are produced in decays 
of unstable particles. In Sec.~\ref{sec:norm} we consider the problem of 
normalization of the neutrino wave packets in the QM framework and show how 
the normalization problem is solved in a natural way in the QFT-based approach. 
In Sec.~\ref{sec:addit} we discuss how one can relax some assumptions usually 
adopted in the QM and QFT approaches. 
Those include the assumption that the maxima of wave packets of all particles 
involved in the neutrino production (or detection) process meet at one 
space-time point, as well as the assumption that the mean momenta of the 
emitted and detected neutrino wave packets coincide. 
We summarize our results and conclude in Sec.~\ref{sec:disc}.
Some technical material is included in Appendices A and B.

%=======================================================================
\section{Review of the QM wave packet formalism
\label{sec:QM}}
%=======================================================================

We start with some generalities that are common to QFT and QM and then 
move on to review the QM wave packet approach to neutrino oscillations. 
We shall use the natural units $\hbar=c=1$ throughout the paper. 

In quantum theory, one-particle states of particles of type $A$ can be 
written as   
\be
|A\rangle =\int\! 
[d p]
\,f_A(\vec{p},
\vec{P})\,|A,\vec{p}
\rangle\,,
\label{eq:state1}
\ee
where $|A,\vec{p}\rangle$ is the one-particle momentum eigenstate  
corresponding to momentum $\vec{p}$ and energy $E_A(\vec{p})$ (for free
particles, $E_A(\vec{p}) = \sqrt{\vec{p}^2 + m_A^2}$, $m_A$ being the mass of
the particle), $f_A(\vec{p},\vec{P})$ is the momentum distribution function
with the mean momentum $\vec{P}$, and we use the shorthand notation 
\be
[dp]\equiv\frac{d^3 p}{(2\pi)^3\sqrt{2E_A(\vec{p})}}\,.
\label{eq:not1}
\ee
For particles with spin, the states $|A\rangle$ and $|A,\vec{p}\rangle$ 
depend also on a spin variable, which we suppress to simplify the notation. 
We will also often omit the second argument of $f_A$ where this cannot 
cause confusion. 

We choose the 
Lorentz invariant normalization condition for the plane 
wave states $|A,\vec{p}\rangle$:\be
\langle 
A,\vec{p}'|A,\vec{p}\rangle= 2E_A(\vec{p})\,(2\pi)^3 
\delta^{(3)}(\vec{p}-\vec{p'})\,. 
\label{eq:norm1}
\ee
The standard normalization of the states $\langle A|A\rangle=1$ then implies 
\be
\int\! \frac{d^3p}{(2\pi)^3}\,|f_A(\vec{p})|^2=1\,. 
\label{eq:norm2} 
\ee
The quantity 
$\sqrt{2E_A(\vec{p})}f_A(\vec{p})$ is actually the momentum-space wave function 
of $A$:  $\sqrt{2E_A(\vec{p})} \times f_A(\vec{p})=\langle\vec{p}|A\rangle$. 
The time dependent wave function is $\sqrt{2E_A(\vec{p})}f_A(\vec{p}) 
e^{-i E_A(\vec{p})t}=\langle\vec{p}|A(t)\rangle$, where $|A(t)\rangle=
e^{-i H t}|A\rangle$ and $H$ is the free Hamiltonian of $A$. The 
coordinate-space wave function $\Psi_A(t,\vec{x})$ is the Lorentz-invariant 
Fourier transform of $\langle\vec{p}|A(t)\rangle$:%
\footnote{
Recall that the Fourier transformation is based on the completeness 
condition for 1-particle momentum eigenstates, which for 
our normalization convention 
reads $\int\frac{d^3 p}{(2\pi)^3 2E_A(\vec{p})}|\vec{p}
\rangle\langle\vec{p}|=\mathbbm{1}_{1 part.}$ (see, e.g., \cite{PeskSchr}, 
eq.~(2.39)). Here the right hand side is the unit operator in the subspace 
of 1-particle states and zero in the rest of the Hilbert space. 
Note that the integration measure $d^3 p/E_A(\vec{p})$ is Lorentz invariant.}
\be
\Psi_A(t,\vec{x})\equiv\langle \vec{x}|A(t)\rangle=
\int\! 
\frac{d^3p}{(2\pi)^3\,2E_A(\vec{p})}\langle\vec{p}|A(t)\rangle
e^{i\vec{p}\vec{x}}\,,
\label{eq:psiA1} 
\ee
or
\be
\Psi_A(t,\vec{x})= \int\! [d p]
\,f_A(\vec{p})e^{-iE_A(\vec{p})t+i\vec{p}\vec{x}}\,.
\label{eq:psiA1a} 
\ee
In the QFT framework, it can be written as 
\be
\Psi_A(x)=\langle 0|\hat{\Psi}_A(x)|A\rangle\,,
\label{eq:psiA2} 
\ee
where $x\equiv (t,\,\vec{x})$ and 
$\hat{\Psi}_A(x)$ is the second-quantized field operator of $A$. Using 
the standard decomposition of the field $\hat{\Psi}_A(x)$ in terms of
production and annihilation operators, one can readily obtain  
(\ref{eq:psiA1}) from (\ref{eq:psiA2}) and (\ref{eq:state1}). Note that 
expressions (\ref{eq:psiA1}) and (\ref{eq:psiA1a}) can describe both 
bound states and propagating wave packets (in the case of bound states or 
particles propagating in a potential, the relation $E_A({\vec{p}}) = 
\sqrt{\vec{p}^2 + m_A^2}$ simply has to be replaced by the proper dispersion
relation). A wave packet 
is obtained when the momentum distribution function $f_A(\vec{p},\vec{P})$ is 
sharply peaked at or close to a nonzero mean momentum $\vec{P}$,%  
\footnote{The peak momentum coincides with the mean momentum for symmetric 
wave packets.}
i.e.\ when the momentum dispersion $\sigma_p$ satisfies $\sigma_p\ll |\vec{P}|$; 
for the rest of this section we will assume this to be the case. The wave 
function~(\ref{eq:psiA1a}) then describes a wave packet whose maximum of 
amplitude is located at $\vec{x}=0$ at time $t=0$. A wave packet that is 
peaked at coordinate $\vec{x}_0$ at time $t_0$ is 
obtained by acting on the state $|A\rangle$ by the space-time translation 
operator $e^{i \hat{P}x_0}$, where $\hat{P}^\mu$ is the 4-momentum operator. 
For the coordinate-space wave function this yields  
\be
\Psi_A(x)=\int\! [d p] \,f_A(\vec{p})e^{-iE_A(\vec{p})(t-t_0)+i\vec{p}
(\vec{x}-\vec{x}_0)} 
\label{eq:psiA3} 
\ee
instead of eq.~(\ref{eq:psiA1a}).

Eqs.~\eqref{eq:psiA1a} and \eqref{eq:psiA3} represent wave packets 
that propagate with the group velocity 
$\vec{v}\equiv\frac{\partial E_A(\vec{p})}{\partial\vec{p}}|_{\vec{p}=\vec{P}}
=\frac{\vec{P}}{E_A(\vec{P})}$ and in general spread with time both in 
the longitudinal direction and in the directions transverse to their mean 
momentum. The spreading is due to the fact that different momentum components 
of the wave packet have slightly different velocities $\vec{p}/E_A(\vec{p})$. 

Let us now consider neutrino oscillations in the framework of the QM wave 
packet formalism, sometimes also called the ``intermediate wave packet'' 
approach. Neutrinos produced or absorbed in charged-current weak interaction 
processes are considered to be flavour eigenstates $\nu_\alpha$ ($\alpha=e,
\mu,\tau)$, which are coherent linear superpositions of mass eigenstates 
$\nu_j$ ($j=1,2, 3)$ with coefficients given by the elements of the leptonic 
mixing matrix $U_{\alpha j}$. The mass eigenstates are represented by the 
corresponding wave packets. 
If a neutrino of flavour $\alpha$ was produced at time $t_P$ at a source 
centered at $\vec{x}_P$, its momentum-space wave function at a time $t> t_P$ is 
\begin{align}
  \sprod{\vec{p}}{\nu_{\alpha P}(t)}
  &= \sum_j U_{\alpha j}^* 
\sqrt{2E_A(\vec{p})}\,f_{jP}(\vec{p},\vec{P}) \, e^{-i E_j(\vec{p})(t - 
t_P) - i\vec{p}\vec{x}_P}\,.
  \label{eq:qm-wp-S}
\end{align}
Here the subscript $P$ shows that the wave packet corresponds to a neutrino 
produced at the source. Note that the index $\alpha$ at $\nu_{\alpha P}(t)$ 
simply indicates that the emitted neutrino was of flavour $\alpha$ at its 
production time $t=t_P$; it is, of course, no longer so for $t>t_P$.   
The shape of the wave packet of the $j$th mass-eigenstate neutrino is given 
by the momentum distribution function $f_{jP}$, which is determined by the 
mechanism and conditions of neutrino production. In the QM framework, however, 
the neutrino production and detection processes are not explicitly taken into 
account; therefore the functions $f_{jP}$ are postulated rather than 
determined, with the corresponding momentum widths estimated from the 
localization properties of the production process. Usually, the wave packets 
are taken to be of the Gaussian form 
\be
f_{jP}(\vec{p},\vec{P})=\left(\frac{2\pi}{\sigma_{pP}^2}\right)^{3/4}
\exp\Big[{-\frac{(\vec{p}-\vec{P})^2}{4\sigma_{pP}^2}}\Big]\,,
\label{eq:Gauss1}
\ee 
where $\sigma_{pP}$ characterizes the momentum uncertainty of the produced 
neutrino state, and similarly for the state of the detected neutrino. The 
advantage of Gaussian wave packets is that they allow most calculations 
to be done analytically (the same is also true for Lorentzian wave
packets, see ref.~\cite{Kopp:2009fa}).

The state of the detected neutrino $\nu_\beta$ 
is described by a wave packet peaked at the detection coordinate $\vec{x}_D$. 
In the momentum-space representation it is given by 
\begin{align}
  \sprod{\vec{p}}{\nu_{\beta D}}
  &= \sum_k U_{\beta k}^* 
\sqrt{2E_A(\vec{p})}\,
f_{kD}(\vec{p},\vec{P}') \, e^{-i \vec{p} 
\vec{x}_D}\,,
\label{eq:qm-wp-D}
\end{align}
where the subscript $D$ stands for detection. The momentum distribution 
functions $f_{kD}$ are governed by the properties of the detection 
process; however, just as for neutrino production, in the QM approach these 
functions are postulated rather than determined. Note that, although the 
assumption $\vec{P}=\vec{P}'$ is adopted in most studies, in general 
there is no reason to expect the mean momenta of the produced and detected 
wave packets to coincide. We will discuss this point in more detail in 
Sec.~\ref{sec:addit}.  

The amplitude for the transition 
$\nu_{\alpha} \to \nu_{\beta}$ is obtained by projecting 
the evolved neutrino production state onto the detection state: 
\be
\mathcal{A}_{\alpha\beta}(T, \vec{L})=\langle \nu_{\beta D}|
\nu_{\alpha P}(t_D)\rangle\,,
\label{eq:amp1}
\ee
where $t_D$ is the detection time, $T \equiv t_D - t_P>0$ and $\vec{L}\equiv 
\vec{x}_D-\vec{x}_P$. Performing the projection in momentum space, 
we obtain from~(\ref{eq:qm-wp-S}) and~(\ref{eq:qm-wp-D})
\footnote{Projection in momentum space will turn out to be convenient for 
our subsequent discussion. The coordinate-space projection 
$\mathcal{A}_{\alpha\beta}(T, \vec{L})= \int\!d^3 x\, 
\sprod{\nu_{\alpha D}}{\vec{x}}\!\sprod{\vec{x}}{\nu_{\alpha P}(t_D)}$ 
yields, of course, the same result.}
\begin{align}
  \mathcal{A}_{\alpha\beta}(T, \vec{L})
    &= \int\!\frac{d^3p}{(2\pi)^3 2E_A(\vec{p})}\, \sprod{\nu_{\beta D}}
{\vec{p}}\!\sprod{\vec{p}}
{\nu_{\alpha P}(t_D)} \nonumber \\
    &= \sum_j U_{\alpha j}^* U_{\beta j}^{}
       \int\!\frac{d^3p}{(2\pi)^3}\, f_{jP}(\vec{p},\vec{P}) \, 
f^*_{jD}(\vec{p},\vec{P}') \,
       e^{-i E_j(\vec{p}) T + i \vec{p} \vec{L}} \,. 
\label{eq:qm-A-1}
\end{align}
For future reference, we shall also write this 
as a superposition of the amplitudes corresponding 
to the contributions of different neutrino mass eigenstates:
\be
\mathcal{A}_{\alpha\beta}(T, \vec{L})= \sum_j U_{\alpha j}^* U_{\beta j}^{}
\mathcal{A}_j(T, \vec{L})
\label{eq:P0}
\ee
with 
\be
\mathcal{A}_j(T, \vec{L})=
       \int\!\frac{d^3p}{(2\pi)^3}\, f_{jP}(\vec{p},\vec{P}) \, 
f^*_{jD}(\vec{p},\vec{P}') \,
       e^{-i E_j(\vec{p}) T + i \vec{p} \vec{L}} \,. 
\label{eq:Aj}
\ee
The oscillation probability is given by the squared modulus of the transition 
amplitude: $P_{\alpha\beta}(T,\vec{L})\equiv P(\nu_\alpha\to\nu_\beta, T,
\vec{L})=|{\cal A}_{\alpha\beta}(T,\vec{L})|^2$. Since in most experiments the 
neutrino emission and detection times are not measured, the standard procedure 
is then to integrate $P(\nu_\alpha\to\nu_\beta, T,\vec{L})$ over $T$. 
This gives 
\be
P(\nu_\alpha\to\nu_\beta,\vec{L})\equiv P_{\alpha\beta}(\vec{L})=
\int dT\, |{\cal A}_{\alpha\beta}(T,\vec{L})|^2\,.
\label{eq:P1}
\ee 
Substituting here the transition amplitude (\ref{eq:qm-A-1}) 
yields, up to a normalization factor, the standard probability of neutrino 
oscillations in vacuum provided that all decoherence effects are negligible. 
The normalization factor can then be fixed by requiring that the oscillation 
probability satisfy the unitarity condition (see 
Sec.~\ref{sec:norm} for a more detailed discussion). 

\section{Neutrino oscillations in QFT \label{sec:QFT}}

\begin{figure}
  \begin{center}
    \includegraphics{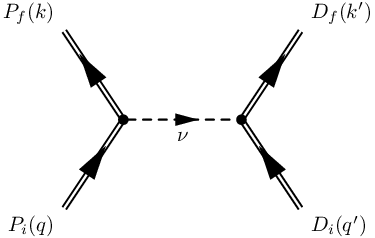}
  \end{center}
\caption{Feynman diagram describing neutrino production, propagation 
           and detection as a single process.}
  \label{fig:feyn}
\end{figure}

In the QFT approach (which is sometimes also called the ``external wave 
packet'' formalism), neutrino production, propagation, and detection are 
considered as a single process, described by the Feynman diagram shown in
fig.~\ref{fig:feyn}, with the neutrino in the intermediate state. In our 
overview of the QFT formalism we will
mostly follow ref.~\cite{Beuthe1}.  Assume 
that the neutrino production process involves one initial state and one final
state particle (besides the neutrino). Likewise, we will assume that the
detection process involves only one particle besides the neutrino in the
initial state and one particle in the final state. The generalization to the
case of an arbitrary number of particles involved in the neutrino production
and detection processes is straightforward and would just complicate the
formulas without providing further physical insight. All external particles
will be assumed to be on their respective mass shells.%
\footnote{
Since only one particle is assumed to be in the initial state of the 
production process, it must be unstable. This will be of no importance for us 
here because, as was already mentioned, the results are easily generalized to 
the case of an arbitrary number of external particles.   
Possible instability of the parent particle will be taken into account in 
Sec.~\ref{sec:unstable}.}

The states describing the particles accompanying neutrino production and 
detection (``external particles'') can be represented in the    
form (\ref{eq:state1}). For the initial and final states at neutrino 
production we can write
\be
|P_i\rangle =\int\! [d q] 
\,f_{Pi}(\vec{q},\vec{Q})\,|P_i,\vec{q}\rangle\,,\qquad
|P_f\rangle =\int\! [d k]
\,f_{Pf}(\vec{k},\vec{K})\,|P_f,\vec{k}\rangle\,,
\label{eq:state2}
\ee
and similarly for the states accompanying neutrino detection: 
\be
|D_i\rangle =\int\! [d q']
\,f_{Di}(\vec{q}',\vec{Q}')\,|D_i,\vec{q}'\rangle\,,\qquad
|D_f\rangle =\int\! [dk']
\,f_{Df}(\vec{k}',\vec{K}')\,|D_f,\vec{k}'\rangle\,.
\label{eq:state3}
\ee
We assume these states to fulfill the normalization condition \eqref{eq:norm2}.
Some (or all) of the mean momenta of the external particles
$\vec{Q}$, $\vec{K}$, $\vec{Q}'$ and $\vec{K}'$ may vanish, i.e.\ the states 
in eqs.~(\ref{eq:state2}) and~(\ref{eq:state3}) can describe bound states 
at rest as well as wave packets. 

%%%%%%%%%%%%%%%%%%%%%%%%%%%%%%%%%%%%%%%%%%%%%%%%%%%%%%%%%%%%%%%%%%%%%
%%%%%%%%%%%%%%%%%%%%%%%%%%%%%%%%%%%%%%%%%%%%%%%%%%%%%%%%%%%%%%%%%%%%%

The amplitude of the neutrino production - propagation - detection process is 
given by the matrix element
\be
i{\cal A}_{\alpha\beta}=\langle P_f \,D_f|\hat{T}\exp\Big[-i\int \!d^4x 
\,{\cal H}_I(x)\Big]-\mathbbm{1}|P_i \,D_i\rangle\,,
\label{eq:amp2}
\ee
where $\hat{T}$ is the time ordering operator and ${\cal H}_I(x)$ is the 
charged-current weak interaction Hamiltonian.%
\footnote{We consider neutrino production and detection at energies well 
below the $W$-boson mass, so that ${\cal H}_I$ is the effective 4-fermion 
Hamiltonian of weak interactions.} 
Note that no neutrino flavour eigenstates have to be introduced in the QFT 
framework, and the indices $\alpha$ and $\beta$ simply refer here to the 
flavour of the charged leptons participating in the production and detection
processes.   

{}From eq.~(\ref{eq:amp2}) it is easy to calculate the transition amplitude 
in the lowest nontrivial (i.e.\ second) order in ${\cal H}_I$ using the 
standard QFT methods. The resulting expression corresponds to the Feynman 
diagram of fig.~\ref{fig:feyn} and can be written as 
\bea
i{\cal A}_{\alpha\beta}&=&\sum_j U_{\alpha j}^* U_{\beta j}
\int\! [dq]
\,f_{Pi}(\vec{q},\vec{Q}) \int\! [dk]
\,f^*_{Pf}(\vec{k},\vec{K}) 
\nonumber \\
& &
\times \int\! [dq']
\, f_{Di}(\vec{q}',\vec{Q}') \int\! [dk']
\,f^*_{Df}(\vec{k}',\vec{K}')
\,i{\cal A}^{p.w.}_j(q,k;q',k')\,.
\label{eq:amp3}  
\eea
Here the sum runs over all intermediate states (i.e.\ different neutrino 
mass eigenstates), and the quantity ${\cal A}^{p.w.}_j(q,k;q',k')$ is the 
plane-wave amplitude of the process with the $j$th neutrino mass eigenstate 
propagating between the source and the detector:
\begin{align}
i{\cal A}^{p.w.}_j(q,k;q',k')=&
\int d^4 x_1 \!\int d^4 x_2 \,
\tilde{M}_D(q',k')\, e^{-i(q'-k')(x_2-x_D)}
\nonumber \\
& \times
i\!\int\!\frac{d^4 p}{(2\pi)^4} \frac{\slashed{p}+m_j}
{p^2-m_j^2+i\epsilon}\, e^{-i p (x_2-x_1)}
\cdot \tilde{M}_P(q,k)\, e^{-i(q-k)(x_1-x_P)} \,. 
\label{eq:amp4}
\end{align}
Here $x_1$ and $x_2$ are the 4-coordinates of the neutrino production 
and detection points,
the quantities $\tilde{M}_P(q,k)$ and $\tilde{M}_D(q',k')$ are the plane-wave
amplitudes of the processes $P_i\to P_f+\nu_j$ and $D_i+\nu_j\to D_f$,
respectively, with the neutrino spinors $\bar{u}_j(p,s)$ and $u_j(p,s)$ 
excluded. The choice of the 4-coordinate dependent phase factors corresponds 
to the assumption that the peaks of the wave packets of particles involved in 
the production process are all located at $\vec{x}_1=\vec{x}_P$ at the time 
$t_1=t_P$, whereas for the detection process the corresponding peaks are all 
situated at $\vec{x}_2=\vec{x}_D$ at the time $t_2=t_D$ (we will discuss in 
Sec.~\ref{sec:addit} how this assumption can be relaxed). The integral in the 
second line of eq.~(\ref{eq:amp4}) gives the coordinate-space propagator of 
the $j$th neutrino mass eigenstate. 

It is convenient to switch to shifted 4-coordinate variables $x_1'$, $x_2'$,
defined according to $x_1=x_P+x_1'$, $x_2=x_D+x_2'$. Taking into account that 
$\slashed{p}+m_j=\sum_s u_j(p,s)\bar{u}_j(p,s)$, one can then rewrite 
eq.~(\ref{eq:amp4}) as 
\begin{align}
i{\cal A}^{p.w.}_j(q,k;q',k')=
i\! \int\!\frac{d^4 p}{(2\pi)^4} \frac{e^{-i p (x_D-x_P)}}{p^2-m_j^2+i\epsilon}
\int \!d^4 & x_1'\,\sqrt{2 p_0}\, M_{jP}(q,k)\,e^{-i(q-k-p)x_1'} 
\nonumber \\
\times 
\int\! d^4 & x_2'\,\sqrt{2 p_0}\,M_{jD}(q',k')\,
e^{-i(q'+p-k')x_2'} \,,
\label{eq:amp5}
\end{align}
where
\be
M_{jP}(q,k) \equiv \frac{\bar{u}_{jL}(p)}{\sqrt{2 p_0}}\tilde{M}_P(q,k)\,\qquad 
\mbox{and}\qquad M_{jD}(q',k')\equiv\tilde{M}_D(q',k')\frac{u_{jL}(p)}
{\sqrt{2 p_0}} 
\label{eq:M}
\ee
are the full amplitudes 
(with the neutrino spinors included) of the processes 
$P_i\to P_f+\nu_j$ and $D_i+\nu_j\to D_f$, respectively, and we have taken 
into account that the matrix elements $M_{jP}(q,k)$ and $M_{jD}(q',k')$ 
involve the left-handed chirality projection, so that only the left-handed 
spinors $u_{jL}(p)$ and $\bar{u}_{jL}(p)$ contribute to the sum over the 
neutrino spin variable~$s$.   
 
Substituting (\ref{eq:amp5}) into eq.~(\ref{eq:amp3}), we finally obtain  
\be
i{\cal A}_{\alpha\beta}=i \sum_j U_{\alpha j}^* U_{\beta j}
\int\!\frac{d^4 p}{(2\pi)^4}\,\Phi_{jP}(p^0, \vec{p}) \Phi_{jD}(p^0, \vec{p})\,
\frac{\,2 p_0\,e^{-i p^0 T + i \vec{p} \vec{L}}}{p^2-m_j^2+i\epsilon}\,.
\label{eq:amp6}
\ee
Here the so-called overlap functions $\Phi_{jP}(p^0, \vec{p})$ and 
$\Phi_{jD}(p^0, \vec{p})$ are defined as  
\begin{align}
\Phi_{jP}(p^0, \vec{p})\,=&\int d^4 x_1' e^{i p x_1'}\int\!
[d q]\int\! [dk]
\,f_{Pi}(\vec{q},\vec{Q})\,f^*_{Pf}(\vec{k},\vec{K}) \,
e^{-i(q-k)x_1'}M_{jP}(q,k)\,, 
\label{eq:psi} \\
\Phi_{jD}(p^0, \vec{p})\,=&\int d^4 x_2' e^{-i p x_2'}\int\!
[dq'] \int\! [dk']
\,f_{Di}(\vec{q}',\vec{Q}')\,f_{Df}^*(\vec{k}',\vec{K}') \,e^{-i(q'-k')x_2'}
M_{jD}(q',k')\,.
\nonumber
\end{align}
Note that they are independent of $x_P$ and $x_D$. Expressions (\ref{eq:amp6}) 
and (\ref{eq:psi}) are the main results of the QFT-based approach to neutrino 
oscillations~\cite{Giunti:2002xg,Beuthe1}. 

\section{Comparing the QM and QFT approaches to neutrino oscillations 
\label{sec:QMQFT}}

Let us now compare the results of the QM and QFT approaches to neutrino 
oscillations. Consider first the transition amplitude~(\ref{eq:amp6}) 
obtained in the QFT formalism. The integration over the neutrino 4-momentum 
in this expression can be done in different order. Here it will be more 
convenient for us to integrate first over  
$p^0$ and then over $\vec{p}$ (the opposite order will be used in 
Sec.~\ref{sec:norm}). Since the distance $L$ between the neutrino source and 
detector is macroscopic,
the phase factor in the integrand of eq.~(\ref{eq:amp6}) undergoes fast 
oscillations and the integral is strongly suppressed except when the 
intermediate neutrino is on the mass shell.  
Thus, the dominant contribution to the integral is given by the residue at 
the pole of the neutrino propagator at $p^0=E_j(\vec{p})-i\epsilon$,% 
\footnote{The contribution of the pole at $p^0=-E_j(\vec{p})+i\epsilon$ is 
strongly suppressed due to an approximate conservation of mean energies at 
production and the fact that $E_{Pi}(\vec{Q})-E_{Pf}(\vec{K})>0$. 
}
where 
\be
E_j(\vec{p}) = \sqrt{\vec{p}^2+m_j^2}\,.
\label{eq:Ej}
\ee
Eq.~(\ref{eq:amp6}) can therefore be rewritten as  
\be
i{\cal A}_{\alpha\beta}= \Theta(T) \sum_j U_{\alpha j}^* U_{\beta j}
\int\!\frac{d^3 p}{(2\pi)^3}\,
\Phi_{jP}(E_j(\vec{p}), \vec{p}) \Phi_{jD}(E_j(\vec{p}), \vec{p})
\,e^{-i E_j(\vec{p}) T + i \vec{p} \vec{L}}\,.
\label{eq:amp7}
\ee
where $\Theta(x)$ is the Heaviside step function.

%-----------------------------------------------------------------------
\subsection{Deriving neutrino wave packets in the QFT-based approach
%-----------------------------------------------------------------------
\label{sec:qft-wp}}

Let us now compare eqs.~(\ref{eq:amp7}) and~(\ref{eq:qm-A-1}). We see that 
the two equations are of the same form and actually coincide if we identify 
the QM wave packets as 
\be
f_{jP}(\vec{p})=\Phi_{jP}(E_j(\vec{p}),\vec{p})\,,
\qquad
f_{jD}(\vec{p})=\Phi^*_{jD}(E_j(\vec{p}),\vec{p})\,,
\label{eq:ident1}
\ee
where the functions $\Phi_{jP}$ and $\Phi_{jD}$ were defined in 
eq.~(\ref{eq:psi}). 

The obtained result can be easily understood. Indeed, as follows from 
the definition of $\Phi_{jP}(p^0,\vec{p})$,  
for $p^0=E_j(\vec{p})$ (i.e.\ on the mass shell of $\nu_j$) this quantity is 
the probability amplitude of 
the production process in which the $j$th mass eigenstate neutrino is emitted 
with momentum $\vec{p}$; but this is nothing but the momentum distribution 
function of the produced neutrino, i.e.\ the momentum-state wave packet 
$f_{jP}(\vec{p})$. A similar argument applies to the neutrino detection 
process and $f_{jD}(\vec{p})$. The wave packets $f_{jP}(\vec{p})$ and 
$f_{jD}(\vec{p})$ in eq.~(\ref{eq:ident1}) are not normalized according to 
(\ref{eq:norm2}), though they can be easily modified to satisfy this 
condition. However, as we shall see in Sec.~\ref{sec:norm}, this is not 
necessary and actually would be misleading.   

An alternative method of deriving 
neutrino wave packets in the QFT framework, based on the S-matrix 
approach, was suggested in~\cite{Giunti:2002xg}; the obtained results are 
equivalent to those in eqs.~(\ref{eq:ident1}) and (\ref{eq:psi}). 

Let us now consider the wave packet describing the produced neutrino 
state in more detail (the state of the detected neutrino can be studied 
quite analogously). 
According to (\ref{eq:ident1}), the momentum distribution function 
$f_{jP}(\vec{p})$ characterizing the state of the emitted neutrino of mass 
$m_j$ is essentially given by the on-shell function $\Phi_{jP}(E_j(\vec{p}),
\vec{p})$. Since the matrix element $M_{jP}(q,k)$ is a smooth function of the 
on-shell 4-momenta $p$ and $q$, whereas the wave packets of the external 
states are assumed to be sharply peaked at or near the corresponding mean 
momenta, one can replace $M_{jP}$ by its value at the mean momenta and pull it out of 
the integral. Eqs.~(\ref{eq:psi}) and (\ref{eq:ident1}) then yield  
\be
f_{jP}(\vec{p})\simeq 
M_{jP}(Q,K) \int d^4 x\, e^{i E_j(\vec{p}) t-i\vec{p}\vec{x}}\!\int\!
[d q] \int\! [d k]
\,f_{Pi}(\vec{q},\vec{Q})\,f_{Pf}^*(\vec{k},\vec{K}) \,e^{-i(q-k)x}\,, 
\label{eq:wp-new-1}
\ee
where the 4-momenta $Q$ and $K$ are defined as 
\be
Q=(E_{Pi}(\vec{Q}),\,\vec{Q})\,,\qquad  
K=(E_{Pf}(\vec{K}),\,\vec{K})\,.
\label{eq:QK}
\ee 
From eq.~(\ref{eq:wp-new-1}) (or eqs.~(\ref{eq:psi}) and (\ref{eq:ident1})) 
one can draw some important conclusions about the properties of the 
neutrino momentum distribution functions $f_{jp}(\vec{p})$ which determine 
the emitted neutrino wave packets: 
\begin{itemize}
\item 
Since the quantities $f_{Pi}(\vec{q},\vec{Q})$ and $f_{Pf}(\vec{k},\vec{K})$ 
depend only on the properties of the external particles, and the 
$j$-dependence of the matrix elements $M_{jP}(p,q)$ comes through the 
on-shell neutrino spinor factors 
$[(2p^0)^{-1/2}u_j(p,s)]_{p^0=E_j(\vec{p})}$, 
which depend on $j$ only through the neutrino energy,
the functions $f_{jP}(\vec{p})$ depend on the index $j$ solely through the 
neutrino energy $E_j(\vec{p})$. This, in particular, means that for 
ultra-relativistic or quasi-degenerate in mass neutrinos the momentum 
distribution functions of all neutrino mass eigenstates are practically 
the same (provided that their energy differences $|E_j-E_k|$ are small 
compared to the energy uncertainty $\sigma_{eP}$).

\item
Because the integral over the 3-coordinate $\vec{x}$ in 
eq.~(\ref{eq:wp-new-1}) yields $\delta^{(3)}(\vec{q}-\vec{k}-\vec{p})$, and 
the momentum distribution functions $f_{Pi}(\vec{q},\vec{Q})$ and 
$f_{Pf}(\vec{k},\vec{K})$ 
are sharply peaked at or near their respective mean momenta $\vec{Q}$ and 
$\vec{K}$, the neutrino momentum distribution functions $f_{jP}(\vec{p})$ are 
sharply peaked at or close to the momentum $\vec{P}\equiv \vec{Q}-\vec{K}$, 
with the width of the peak $\sigma_{pP}$ dominated by the largest between the 
momentum uncertainties of the states of $P_i$ and $P_f$. 

\end{itemize}

Taking into account eq.~(\ref{eq:psiA1}), eq.~(\ref{eq:wp-new-1}) can be 
rewritten as
\be
f_{jP}(\vec{p})\simeq M_{jP}(Q,K) \int d^4 x \,
e^{i p x}\, 
\Psi_{P_i}(x)\Psi_{P_f}^*(x)\Big|_{p^0=E_j(\vec{p})}\,. 
\label{eq:wp-new-2}
\ee
Thus, the momentum distribution function that determines the wave packet of 
the emitted neutrino is essentially the 4-dimensional Fourier transform of 
the product of the coordinate-space wave functions of the external particles 
participating in the neutrino production process, 
taken under the condition that the components of the neutrino 4-momentum 
are on the mass shell. Eq.~(\ref{eq:wp-new-2}) can be readily 
generalized to the case when more than two external particles 
participate in the neutrino production process: the expression 
$\Psi_{P_i}(t,\vec{x})\Psi_{P_f}^*(t,\vec{x})$ in the integrand 
of~(\ref{eq:wp-new-2}) should simply be replaced by the product of the wave 
functions of all particles in the initial state of the production process and 
of complex conjugates of the wave functions of all particles in the final 
state (except the neutrino).

The neutrino wave packet in coordinate space $\psi_{jP}(x)$ is 
obtained from eq.~(\ref{eq:wp-new-2}) by performing the Fourier 
transformation over the 3-momentum variable $\vec{p}$ according to the 
transformation law (\ref{eq:psiA1a}), which gives 
\begin{align}
\psi_{jP}(x)\simeq &
\;\bar{u}_j(P,s)\tilde{M}_{jP}(Q,K)\,\int d^4 x'\, 
\Psi_{P_i}(x')\Psi_{P_f}^*(x') \nonumber \\ & \times 
\frac{\Theta(t-t')}{|\vec{x}-\vec{x}'|}\frac{-i}{2(2\pi)^2}\int_0^\infty 
\frac{dp\, p}{\sqrt{p^2+m_j^2}}\,e^{-i \sqrt{p^2+m_j^2}(t-t')}
\big[e^{ip|\vec{x}-\vec{x}'|}-e^{-ip|\vec{x}-\vec{x}'|}\big]\,.
\label{eq:wp-new-3}
\end{align}
Here $p\equiv|\vec{p}|$ and we have used eq.~(\ref{eq:M}) to extract the 
$p$-dependent factor $(2 p^0)^{-1/2}=2^{-1/2}(\vec{p}^2+m_j^2)^{-1/4}$ from 
$M_{jP}(Q,K)$. The integral over $p$ in eq.~(\ref{eq:wp-new-3}) can be 
expressed in terms of the modified Bessel function $K_1$ \cite{Gradshtein}, 
giving 
\be
\psi_{jP}(x)\simeq \frac{\bar{u}_j(P,s)\tilde{M}_{jP}(Q,K)}
{(2\pi)^2}\!\int d^4 x'\, 
\Psi_{P_i}(x')\Psi_{P_f}^*(x') 
\,\frac{m_j\, \Theta(t-t')}{\sqrt{-(x-x')^2}}\,K_1(m_j\,\sqrt{-(x-x')^2}\,),
\label{eq:wp-new-4}
\ee
where $(x-x')^2\equiv(t-t')^2-(\vec{x}-\vec{x}')^2$. 
The integral in the second line of eq.~(\ref{eq:wp-new-3}) is greatly 
simplified in the limit of vanishing neutrino mass:    
\be
\psi_{jP}(x)\simeq \bar{u}_j(P,s)\,\tilde{M}_{jP}(Q,K)
\frac{-1}{(2\pi)^2}\int d^4 x'\, 
\Psi_{P_i}(x')\Psi_{P_f}^*(x')
\frac{\Theta(t-t')}
{(x-x')^2}\,.
\label{eq:wp-new-5}
\ee
Note, however, that in this limit all neutrino species travel with the same 
speed and therefore the wave functions in eq.~(\ref{eq:wp-new-5}) cannot 
describe decoherence due to the separation of wave packets. In order to take 
possible wave packet separation effects into account the more accurate 
expression (\ref{eq:wp-new-4}) has to be used. Alternatively, one can employ 
the momentum-representation wave function~(\ref{eq:wp-new-2}). 

Expression (\ref{eq:wp-new-4}) for the wave function of the produced neutrino 
state $\Psi_{jP}(t,\vec{x})$ allows a simple interpretation.  Note that 
$\frac{1}{(2\pi)^2}\,m_j \Theta(t-t')K_1(m_j\sqrt{-(x-x')^2})/
{\sqrt{-(x-x')^2}}\,$ is the scalar retarded propagator in the coordinate 
representation. 
Therefore the neutrino wave packet (\ref{eq:wp-new-4}) is essentially the 
convolution of the neutrino source (the role of which is played by the 
neutrino production amplitude $\Psi_{P_f}^*(x)M_{jP}\Psi_{P_i}(x)$) with the 
retarded neutrino propagator, in  full agreement with the well known result 
of QFT. Note that only the scalar part of the propagator contributes to 
$\Psi_{jP}(x)$; this is because the coordinate space and momentum space 
neutrino wave functions $\Psi_{jP}(x)$ and $f_{jP}(\vec{p})$ are scalars in our 
formalism. The spinor factors are included in the matrix elements $M_{jP}$ and 
$M_{jD}$ (note that these quantities are also scalar, whereas the amputated 
matrix elements $\tilde{M}_{jP}$ and $\tilde{M}_{jD}$, i.e.\ those  
with the neutrino spinors removed, have spinorial indices).  

In our discussion of the wave packets of the emitted neutrino states, we 
were assuming that the momentum distribution functions of all the external 
particles accompanying neutrino production are known. This implies, in 
particular, that all particles in the final state of the production process 
are ``measured'', either by direct detection or through their interaction 
with the medium in the process of neutrino production. It is quite possible, 
however, that some of the particles accompanying neutrino production escape
undetected; this is, e.g., the case for atmospheric or accelerator neutrinos 
born in the process $\pi^\pm \to \mu^\pm+\nu_\mu(\bar{\nu}_\mu)$, in which 
the final state muon is normally not detected. It is also possible that some 
of the particles accompanying neutrino detection are ``unmeasured''. 
How can one determine the neutrino wave packets in those cases?  

To answer this question, let us recall that 
the momentum uncertainty $\sigma_{pP}$ 
characterizing the emitted neutrino depends in general on the momentum 
uncertainties of all the external particles at neutrino production and is 
dominated by the largest among them (see the discussion after 
eqs.~(\ref{eq:wp-new-1}) and~(\ref{eq:QK})). In particular, in the case of 
Gaussian wave packets, one has~\cite{Giunti:2002xg,Beuthe1}
\be
\sigma_{pP}^2=\sigma_{pPi}^2+\sigma_{pPf}^2\,.
\label{eq:sigmaP}
\ee
For more than two external particles at production, the sum on the right-hand 
side of this relation would contain the contributions of the squared momentum 
uncertainties of all these particles. Now, if a particle goes ``unmeasured'' 
in the neutrino production process, its momentum uncertainty cannot affect the 
momentum uncertainty of the emitted neutrino state and therefore can be 
neglected. To put it differently, undetected particles are completely 
delocalized, and therefore, according to Heisenberg's uncertainty relation, 
have vanishing momentum uncertainty. 
This means that undetected particles can be represented by 
states of definite momenta, i.e.\ by plane waves. If, for example, the particle 
$P_f$ at production is undetected, one has to replace in 
eq.~(\ref{eq:wp-new-1}) the momentum distribution function $f_{Pf}(\vec{k},
\vec{K})$ by $[(2\pi)^3/\sqrt{V}]\delta^{(3)} (\vec{k}-\vec{K})$ where $V$ is 
the normalization volume, and in eq.~(\ref{eq:wp-new-2}) the coordinate-space 
wave function $\Psi_{Pf}(x)$ by 
$e^{-i K x}/\sqrt{2E_{Pf}(\vec{K}) V}$, 
with eqs.~(\ref{eq:wp-new-3}) - (\ref{eq:wp-new-5}) modified accordingly.   
The mean momentum of the neutrino state depends, of course, on the momentum of 
the undetected particle; if the latter can take values in some range, the same 
will be true for the mean momentum of the emitted neutrino state. In this case 
the flux of emitted neutrinos will be characterized by a continuous spectrum. 

In most of our discussion in this subsection we concentrated on the wave 
packets of the produced neutrino states. Our consideration, however, 
applies practically without changes to the detected neutrino states; the 
corresponding formulas can be obtained from eqs.~(\ref{eq:wp-new-1}) and 
(\ref{eq:wp-new-2})-(\ref{eq:wp-new-5}) by replacing, where appropriate,
$e^{i p x} \to e^{-i p x}$, $\Psi_{Pi}\to \Psi_{Di}$ and 
$\Psi_{Pf}\to \Psi_{Df}$.

%-----------------------------------------------------------------------
\subsection{General properties of neutrino wave packets
\label{sec:wp-gen}}
%-----------------------------------------------------------------------

We have already considered some of the general properties of the neutrino wave 
packets in the previous subsection. In particular, we have found that the 
momentum distribution functions $f_{jP}(\vec{p})$ of mass-eigenstate neutrinos 
$\nu_j$ depend on the index $j$ only through the neutrino energy 
$E_j(\vec{p})$, and that the functions $f_{jP}(\vec{p})$ are sharply 
peaked at or near the momentum $\vec{P}=\vec{Q}-\vec{K}$, with the width 
of the peak dominated by the largest between the widths of the functions 
$f_{Pi}$ and $f_{Pf}$. 
Further insight into the general properties of the neutrino wave packets 
can be gained by comparing expressions (\ref{eq:psi}) with their plane-wave 
limits. If the external particles were described by plane waves, the 
quantities $\Phi_{jP}(E_j(\vec{p}),\vec{p})$ and $\Phi_{jD}(E_j(\vec{p}),
\vec{p})$, which determine the neutrino wave packets, would be just equal to 
the matrix elements of the neutrino production or detection processes 
divided by the factor $\sqrt{2E V}$ for each external particle and  
multiplied, correspondingly, by $(2\pi)^4\delta^{(4)}(q-k\mp p)$. The latter 
factors represent energy-momentum conservation at the production and 
detection vertices. As follows from (\ref{eq:psi}), in the case when the 
external particles are described by wave packets, the quantities 
$\Phi_{jP}(E_j(\vec{p}),\vec{p})$ and $\Phi_{jD}(E_j(\vec{p}),\vec{p})$
(and therefore the momentum distribution functions of the neutrino wave 
packets) correspond to ``smeared $\delta$-functions'', representing 
approximate conservation of the mean energies and mean momenta of the 
participating particles. 
How exactly this smearing occurs will depend on the form of the wave packets 
of the external particles, and to move ahead one has to specify this 
form. 

A particularly useful and illuminating example of a specific form of the 
external wave packets, and the one most often used in the literature, is the 
case of Gaussian wave packets. 
We will employ this example to illustrate the general properties of the 
neutrino wave packets. 

Let us discuss first the conditions under which an 
arbitrary wave packet can be accurately approximated by a Gaussian one.
For simplicity, we will consider here 1-dimensional wave packets. This is a 
good approximation in the case when the distance between the neutrino source 
and detector is very large compared to their sizes, so that the neutrino 
momentum is practically collinear with $\vec{L}=\vec{x}_D-\vec{x}_P$ (the 
generalization to the 3-dimensional case is straightforward).  
Consider a wave packet described by a momentum distribution function $f(p)$, 
sharply peaked at some value $P_0$ of the momentum. We can write this function 
in the exponential form as  
\be
f(p)=e^{-g(p)}\,,\qquad \mbox{where} \qquad 
g(p)=-\ln[f(p)]\,.
\label{eq:exp}
\ee 
The Gaussian approximation corresponds to the case when the integral over 
$p$ of the function $f(p)$ multiplied by any function of $p$ that is 
smooth in the vicinity of $P_0$ can be evaluated in the saddle point 
approximation. 
Indeed, in this approach one expands the function $g(p)$ around its 
minimum at 
$p=P_0$ and keeps the terms up to and including the quadratic one:
\be
g(p)\simeq g(P_0)+\frac{1}{2}\, g''(P_0) (p-P_0)^2 \,. 
\label{eq:expan}
\ee
This precisely means the wave packet $f(p)$ is approximated by the Gaussian 
one. The validity condition for this approximation is given in terms of the 
derivatives of the function $g$ at $P_0$: 
\be
\frac{1}{4!}\,|g^{(IV)}(P_0)|\ll \frac{1}{2}\,|g''(P_0)|^2\,.
\label{eq:cond1}
\ee
It can be satisfied for a wide range of functions $f(p)$. However, it is 
easy to construct wave packets for which it is not satisfied. Consider, 
e.g.\ a class of wave packets 
\be
f(p)=\frac{C_n}{[(p-P_0)^2+\gamma^2]^n}\,
\label{eq:class}
\ee
with integer $n$ and $C_n$ a 
constant, which can be found from 
the normalization condition for $f(p)$. It is easy to check that 
condition~(\ref{eq:cond1}) is equivalent to $1/4n \ll 1$. Thus, the momentum 
distribution functions (\ref{eq:class}) can be accurately approximated by the 
Gaussian ones only when $n\gg 1$. This condition, in particular, is not 
satisfied for Lorentzian wave packets.

%-----------------------------------------------------------------------
\subsection{Matching the QFT and QM neutrino wave packets
\label{sec:wp-match}}
%-----------------------------------------------------------------------

Let us now discuss how one can match the QFT and QM wave packets of neutrinos.  
Using the case of Gaussian wave packets as an example, we shall find out 
how the effective parameters describing the QM wave packets can be expressed 
in terms of the more fundamental parameters entering into the QFT approach.  

We start by introducing some notation (we mostly follow \cite{Giunti:2002xg,
Beuthe1} here). 
The coordinate uncertainty $\sigma_{xPi}$ characterizing the wave function of
the initial state particle $P_i$ in the neutrino production process is related 
to its momentum uncertainty $\sigma_{pPi}$ by 
\be
\sigma_{xPi}\, \sigma_{pPi}=\frac{1}{2}\,,
\label{eq:Heisenb}
\ee	
and similarly for all other external particles. One can also introduce the 
effective coordinate uncertainty of the production process $\sigma_{xP}$, 
which is connected to the effective momentum uncertainty of this process 
$\sigma_{pP}$ defined in eq.~(\ref{eq:sigmaP}) by a relation similar 
to (\ref{eq:Heisenb}), or equivalently
\be
\frac{1}{\sigma_{xP}^2}=\frac{1}{\sigma_{xPi}^2}+
\frac{1}{\sigma_{xPf}^2}\,.
\label{eq:sigmaXP}
\ee
This formula has a simple physical interpretation: since the neutrino 
production process requires an overlap of the wave functions of all the 
participating particles, the effective uncertainty of the coordinate of 
the production point is determined by the particle with the smallest 
coordinate uncertainty. 
This is in accord with the already discussed fact 
that the effective momentum uncertainty at production $\sigma_{pP}$, which 
determines the momentum uncertainty of the produced neutrino, is dominated 
by the largest among the momentum uncertainties of all the external particles 
involved in neutrino production. 

Next, we define the effective velocity of the neutrino source $\vec{v}_P$ 
and its effective squared velocity $\Sigma_P$ as 
\be
\vec{v}_P\equiv \sigma_{xP}^2\left(\frac{\vec{v}_{Pi}}{\sigma_{xPi}^2}
+\frac{\vec{v}_{Pf}}{\sigma_{xPf}^2}\right)\,,
\qquad
\Sigma_P\equiv \sigma_{xP}^2\left(\frac{\vec{v}_{Pi}^2}{\sigma_{xPi}^2}
+\frac{\vec{v}_{Pf}^2}{\sigma_{xPf}^2}\right)\,. 
\label{eq:vSigma}
\ee
If $\vec{v}_{Pi} \sim \vec{v}_{Pf}$, they are approximately equal to, 
respectively, the velocity and squared velocity of the particle with the 
smallest coordinate uncertainty. We will also need the quantity $\sigma_{eP}$ 
defined through 
\be
\sigma_{eP}^2=\sigma_{pP}^2 (\Sigma_P-\vec{v}_P^2)
\equiv \sigma_{pP}^2\, \lambda_P \,.
\label{eq:sigmaEP}
\ee
This quantity can be interpreted as the effective energy uncertainty at 
neutrino production~\cite{Beuthe1}. It can be also shown that 
$0\le \lambda_P\le 1$, i.e.\ $0\le \sigma_{eP}\le \sigma_{pP}$.  

We can now discuss the results obtained in the QFT framework in the 
case when the external particles are represented by Gaussian wave 
packets. The function $\Phi_{jP}(E_j(\vec{p}),\vec{p})$, which coincides 
with the momentum 
distribution function of the emitted mass-eigenstate neutrino $\nu_j$, 
can be written as~\cite{Giunti:2002xg,Beuthe1} 
\be 
\Phi_{jP}(E_j(\vec{p}),\,\vec{p}) = N_P\, M_{jP}(Q,K)\,\frac{1}{\sigma_{eP}
\sigma_{pP}^3}\,\exp\big[-g_{P}(E_j(\vec{p}),\,\vec{p})\big]\,, 
\label{eq:genwp} 
\ee 
where
\be
N_P=\frac{\pi^2}{(2\pi \sigma_{xPi}^2)^{3/4} (2\pi \sigma_{xPf}^2)^{3/4}
[2E_{Pi}(\vec{Q})\! \cdot\! 2E_{Pf}(\vec{K})]^{1/2}} 
\label{eq:NP}
\ee
is the normalization factor and 
\be 
g_{P}(E_j(\vec{p}),\,\vec{p})=\frac{(\vec{p}-\vec{P})^2}{4\sigma_{pP}^2}+ 
\frac{[E_j(\vec{p})-E_P-\vec{v}_P(\vec{p}-\vec{P})]^2}{4\sigma_{eP}^2}\,.
\label{eq:g1} 
\ee 
Here 
\be 
\vec{P}\equiv\vec{Q}-\vec{K}\,,\quad 
E_P\equiv E_{Pi}(\vec{Q})-E_{Pf}(\vec{K})\,, 
\label{eq:enmom} 
\ee 
and $E_j(\vec{p})$ was defined in eq.~(\ref{eq:Ej}). Note that in the limit 
when the external particles are represented by plane waves ($\sigma_{pPi}
\to 0$, $\sigma_{pPf}\to 0$),  
the first equation in (\ref{eq:psi}) yields  
\begin{align}
\Phi_{jP}(E_j(\vec{p}),\,\vec{p})= (2\pi)^4 
\delta [E_{Pi}(\vec{Q})-E_{Pf}(\vec{K})-E_j(\vec{P})] \,\delta^{(3)} 
(\vec{Q}-\vec{K}-\vec{P})~~~~~~~\, \nonumber \\
\times
 \frac{M_{jP}(Q,K)}{\sqrt{2E_{Pi}(\vec{Q})V \!
\cdot\! 2E_{Pf}(\vec{K})V}} \,, 
\label{eq:pwlimit} 
\end{align}
as discussed in the previous subsection. From eq.~(\ref{eq:sigmaP}) and the 
fact that $\sigma_{eP}\le \sigma_{pP}$ it follows that in this limit the 
momentum uncertainty $\sigma_{pP}$ and energy uncertainty $\sigma_{eP}$ of the 
produced neutrino state vanish as well; thus, if the external particles are 
described by plane waves, then so is the produced neutrino.  
As follows from eq.~(\ref{eq:pwlimit}), the plane wave limit corresponds 
to exact energy and momentum conservation at production.%
\footnote{Since the neutrino energy and momentum are completely determined 
by those of the external particles in this case, only one neutrino mass
eigenstate can be produced in any given interaction process, and therefore 
no oscillations are possible in the plane wave limit \cite{Beuthe1}.}
This can also be seen from eqs.~(\ref{eq:genwp}) 
and (\ref{eq:g1}):  indeed, in the limit 
$\sigma_{eP},\sigma_{pP}\to 0$ the right hand side of (\ref{eq:genwp}) 
is proportional to  the product of the energy and 
momentum conserving $\delta$-functions.  
For finite values of $\sigma_{pP}$ and $\sigma_{eP}$, eqs.~(\ref{eq:genwp}) 
and~(\ref{eq:g1}) yield Gaussian-type ``smeared delta functions'', i.e.\ 
describe approximate conservation laws for the mean momenta 
and mean energies of the wave packets, for which are responsible, 
respectively, the first term and the second term in~(\ref{eq:g1}). 

Let us now try to cast expressions~(\ref{eq:genwp}) and~(\ref{eq:g1}) into 
the form usually adopted in the QM wave packet approach to neutrino oscillations. 
We want to reduce $\Phi_{jP}(E_j(\vec{p}),\vec{p})$ to 
an expression similar to that in eq.~(\ref{eq:Gauss1}). 
To this end, we expand the neutrino energy around the point $\vec{p}=\vec{P}$ 
and keep terms up to the second order: 
\be
E_j(\vec{p})\simeq E_j+\vec{v}_j (\vec{p}-\vec{P})
+\frac{1}{2E_j}(\delta^{kl}-v_j^k v_j^l)(p-P)^k (p-P)^l\,.
\label{eq:2ndorder}
\ee
Here
\be
E_j\equiv E_j(\vec{P})\,, \qquad \vec{v}_j\equiv
\frac{\partial E_j(\vec{p})}{\partial \vec{p}}\Big|_{\vec{p}=\vec{P}}=
\frac{\vec{P}}{E_j}\,.
\label{eq:Ejj}
\ee
The lower index $j$ corresponds to the neutrino mass eigenstates, 
while the upper indices $k$ and $l$ number the components of the 3-vectors 
$\vec{p}$, $\vec{P}$ and $\vec{v}_j$ (i.e.\ $v_j^k$ is the $k$th 
component of the group velocity of the $j$th neutrino mass eigenstate). 
The function $g_P(E_j(\vec{p}),\vec{p})$ defined in (\ref{eq:g1}) can 
then be written as   
\be
g_P(E_j(\vec{p}),\vec{p})=(p-P)^k\,\alpha^{kl}\,(p-P)^l 
-\beta^k (p-P)^k+\gamma_j\,, 
\label{eq:gP}
\ee
where 
\be
\alpha^{kl}=\frac{1}{4\sigma_{eP}^2}\left[\lambda_P\, \delta^{kl}+(v_j-v_P)^k\, 
(v_j-v_P)^l+\frac{E_j-E_P}{E_j}(\delta^{kl}-v_j^k v_j^l)\right]\,,
\label{eq:alpha}
\ee
\be
\beta^k=-\frac{1}{2\sigma_{eP}^2}(E_j-E_P)(v_j-v_P)^k\,,\qquad 
\gamma_j=\frac{(E_j-E_P)^2}{4\sigma_{eP}^2}\,.
\label{eq:betagamma}
\ee
Let us now try to represent $g_P(E_j(\vec{p}),\vec{p})$ in the form  
\be
g_P(E_j(\vec{p}),\vec{p})=(p-P_{\rm eff})^k\,\alpha^{kl}\,(p-P_{\rm eff})^l 
+\tilde{\gamma}_j 
\label{eq:gP2}
\ee
with 
\be
\vec{P}_{\rm eff}\equiv\vec{P}+\mbox{\boldmath{$\delta$}}\,,
\label{eq:Peff2}
\ee
where the parameters \mbox{\boldmath{$\delta$}} and $\tilde{\gamma}_j$  
are to be determined by comparing eqs.~(\ref{eq:gP2}) and~(\ref{eq:gP}). 
They describe, respectively, a shift of the neutrino mean momentum compared 
to the naive expectation $\langle\vec{p}\rangle=\vec{P}$ and a modification 
of the overall normalization of the neutrino wave function. 
The effective momentum uncertainty characterizing the QM neutrino wave 
packet can be obtained by diagonalizing the matrix $\alpha$. The squared 
uncertainties of the different components of the neutrino momentum are 
given, up to the factor 1/4, by the reciprocals of the eigenvalues of 
the matrix $\alpha$.  In  general, these eigenvalues are different, i.e.\ the 
neutrino momentum uncertainty is anisotropic. This actually means that 
expression (\ref{eq:Gauss1}) for a 3-dimensional Gaussian wave packet is 
oversimplified: its exponent has to be replaced by 
$-[(p^x-P_{\rm eff}^x)^2/4(\sigma_{pP}^x)^2+(p^y-P_{\rm eff}^y)^2/4
(\sigma_{pP}^y)^2 +(p^z-P_{\rm eff}^z)^2/4(\sigma_{pP}^z)^2]$. 

Comparing eqs.~(\ref{eq:gP2}) and~(\ref{eq:gP}), 
one finds that the shift \mbox{\boldmath{$\delta$}} of the neutrino mean 
momentum satisfies the equation  
\be
2\,\alpha^{kl}\,\delta^l=\beta^k\,,
\label{eq:shiftEq}
\ee
whereas the parameter $\tilde{\gamma}_j$ 
is given by
\be
\tilde{\gamma}_j=\gamma_j-\delta^k\,\alpha^{lk}\,\delta^l=\gamma_j-\frac{1}{2}
\delta^k\beta^k\,. 
\label{eq:normtot}
\ee
The full solutions of eqs.~(\ref{eq:shiftEq}) and~(\ref{eq:normtot}) 
are given in Appendix~A; here we present the results obtained in the 
leading order in the small parameter $(E_j-E_P)/E_j$.%
\footnote{That this parameter is indeed small can be readily seen from 
eqs.~(\ref{eq:genwp}) and~(\ref{eq:g1}): the function $\Phi_{jP}$ is 
strongly suppressed unless $|E_j-E_P|\lesssim\max\{\sigma_{eP},\sigma_{pP}\}
\ll E_j$.}
The diagonalization of the matrix $\alpha^{kl}$ gives in this limit
\be
(\sigma_{pP\,{\rm eff}}^x)^2=(\sigma_{pP\,{\rm eff}}^y)^2=\sigma_{pP}^2\,,
\qquad \frac{1}{(\sigma_{pP\,{\rm eff}}^z)^2}=\frac{1}{\sigma_{pP}^2}+
\frac{(\vec{v}_j-\vec{v}_P)^2}{\sigma_{eP}^2}\,,
\label{eq:sigmaPeff3}
\ee
where the $z$ axis was chosen in the direction of $\vec{v}_j-\vec{v}_P$. 
For $\delta^k$ and $\tilde{\gamma}_j$ one finds 
\be
\delta^k=-\frac{(E_j-E_P)(v_{j}-v_{P})^k}{\lambda_P+(\vec{v}_j-\vec{v}_P)^2}\,,
\qquad\qquad \tilde{\gamma}_j = \frac{(E_j-E_P)^2}{4\sigma_{eP}^2}
\frac{\lambda_P}{\lambda_P+(\vec{v}_j-\vec{v}_P)^2}\,.
\label{eq:deltaK2}
\ee
{}From eq.~(\ref{eq:sigmaPeff3}) it follows that the neutrino momentum 
uncertainty in the direction of $\vec{v}_j-\vec{v}_P$ is smaller than those 
in the orthogonal directions. Note that for non-relativistic sources 
($v_P\ll 1$) the direction of $\vec{v}_j-\vec{v}_P$ essentially coincides 
with that of the mean neutrino momentum, and eq.~(\ref{eq:sigmaPeff3}) means 
that the longitudinal uncertainty of the neutrino momentum is smaller 
than the transversal ones. 
To understand this property qualitatively, one can 
imagine the neutrino production region (i.e.\ the region where the wave packets 
of the particles involved in the production process have significant overlap)
to be approximately spherical with radius of order $\sigma_{xP}$. Then, the 
transverse extent of the neutrino wave packet will also be $\mathcal{O}
(\sigma_{xP})$. On the other hand, its longitudinal spread is determined by 
the duration of the production process (i.e.\ the time interval during which 
the wave packets have significant overlap), which is given by $\sim 
\sigma_{xP} / \delta v$, where $\delta v$ is the relative velocity of the two 
external particles.

Summarizing the results of the current subsection, we can see that 
the QM neutrino wave packets can match those obtained in the QFT framework 
if one applies the following changes to the QM results: 
\begin{itemize}
\item 
The momentum uncertainties of the neutrino mass eigenstates are replaced by 
the effective ones, defined in eq.~(\ref{eq:sigmaPeff3}). They are in general 
different in different directions.

\item 
The mean momentum $\vec{P}$ is shifted according to $\vec{P}\to 
\vec{P}_{\rm eff}=\vec{P}+\mbox{\boldmath{$\delta$}}$, where the components 
of \mbox{\boldmath{$\delta$}} are given in eq.~(\ref{eq:deltaK2}). 

\item 
The wave packet of each neutrino mass eigenstate gets an extra factor 
$N_j=\exp[-\tilde{\gamma}_j]$, where $\tilde{\gamma}_j$ is given 
in~(\ref{eq:deltaK2}). 

\end{itemize} 
From the last point one can see that if the differences of the energies of 
different neutrino mass eigenstates are small compared to the energy 
uncertainty $\sigma_{eP}$, 
\be
|E_i-E_j|\simeq \frac{\Delta m_{ij}^2}{2\bar{E}}\,\ll\, \sigma_{eP}\,, 
\label{eq:cond2}
\ee
the additional factors $N_j$ are essentially the same for the wave 
functions of all neutrino mass eigenstates and can be included in their 
common normalization factor. If, on the contrary, this condition is 
violated, the coherence of the emission of different neutrino mass 
eigenstates will be lost \cite{AS}. 

As follows from eq.~(\ref{eq:sigmaPeff3}), the effective momentum 
uncertainties that should be used to describe the wave packets 
of emitted neutrinos in the QM formalism are not just equal to the true 
momentum uncertainty at production $\sigma_{pP}$, as naively expected, but 
also depend on the energy uncertainty $\sigma_{eP}$, which is an independent 
parameter, as well as on the neutrino velocity $\vec{v}_j$ and the effective 
velocity of the neutrino production region $\vec{v}_P$. Except for $\vec{v}_j
\simeq \vec{v}_P$, the momentum uncertainty along the direction of 
$\vec{v}_j-\vec{v}_P$ is dominated by the smaller between $\sigma_{pP}$ 
and $\sigma_{eP}$, which turns out to be $\sigma_{eP}$. This 
is related to the fact that neutrinos propagate over macroscopic 
distances and therefore are on their mass shell, which
enforces the relation $E_j(\vec{p}) \sigma_{eP} \simeq |\vec{p}| 
\sigma_{pP\,{\rm eff}}$ (see Sec.~5.2 of ref.~\cite{AS}).
In  the limit $\sigma_{eP},\vec{v}_P\to 0$, which corresponds to a stationary 
neutrino source approximation \cite{Beuthe1}, the effective longitudinal 
momentum uncertainty $\sigma_{pP\,{\rm eff}}^z$ vanishes, even though the  
true momentum uncertainty $\sigma_{pP}$ is nonzero. This implies an 
infinite coherence length, in accordance with the well known result for the 
stationary case \cite{Kiers,Grimus:1996av}. It also confirms the expectation
that wave packets of \MB\ neutrinos, which are emitted in a quasi-stationary
process, have a very large spatial extent~\cite{AKL1,Akhmedov:2008zz}.

We have discussed here the matching of the QFT and QM wave packets of the 
produced neutrino states; for the wave packets of the detected states 
the consideration is completely analogous. 

\subsection{The case of an unstable neutrino source
\label{sec:unstable}}

Let us now consider the situation when neutrinos are produced in decays 
of unstable particles. Once again, we will 
assume that the external particles are described by Gaussian wave packets. 
Compared to the standard formalism that leads to eqs.~(\ref{eq:genwp}) 
and~(\ref{eq:g1}), one now has to introduce the following modifications:

\begin{itemize}
\item[1.]
The energy of the parent particle $P_i$ acquires an imaginary part, i.e.\ one 
has to replace $E_{Pi}(\vec{q})\to E_{Pi}(\vec{q})-i\Gamma/2$, where 
$\Gamma=[m_{Pi}/E_{Pi}(\vec{q})]\Gamma_0 \simeq [m_{Pi}/E_{Pi}(\vec{Q})]
\Gamma_0$, $\Gamma_0$ being the rest-frame decay width of $P_i$. This amounts 
to replacing the energy difference $E_P$ defined in~(\ref{eq:enmom}) according 
to 
$E_P\to E_P-i\Gamma/2$. 

\item[2.]

The integration over time in the formula for $\Phi_{jP}$ in 
eq.~(\ref{eq:psi}) 
now has to be performed from 0 to $\infty$ rather than from $-\infty$ to 
$\infty$ (assuming that $t=0$ is the production time of the parent particle 
$P_i$). 

\end{itemize}
As a result of these modifications, eqs.~(\ref{eq:genwp}), (\ref{eq:g1}) get 
replaced by 
\be
\Phi_{jP}(E_j(\vec{p}),\,\vec{p}) =
N_P\,M_{jP}(Q,K)\,\frac{1}{\sqrt{\pi} \sigma_{pP}^3}\,
\exp\big[-\frac{(\vec{p}-\vec{P})^2}{4\sigma_{pP}^2} 
\big]\,I_1\,,
\label{eq:genwp2} 
\ee
where 
\be
I_1\equiv 
\int_0^\infty e^{-a(t-t_P)^2+i b (t-t_P)-\frac{\Gamma}{2}t}dt
=e^{-\frac{\Gamma}{2}t_P}
\int_{-t_P}^\infty e^{-a t^2+i \tilde{b} t}dt\,,
\label{eq:I1}
\ee
with 
\be
a\equiv \sigma_{eP}^2,\,\qquad 
b\equiv E_j(\vec{p})-E_P-\vec{v}_P(\vec{p}-\vec{P}) \,,\qquad
\tilde{b} \equiv b+i\Gamma/2\,, 
\label{eq:ab}
\ee
and $t_P \ge 0$ being the time of maximum wave packet overlap in the 
production region (see Sec.~\ref{sec:QFT}).
The $\vec{p}$-dependent 
Gaussian factor in eq.~(\ref{eq:genwp2}) describes, as before, an approximate 
conservation of mean momenta at neutrino production, whereas the factor 
$I_1$ is responsible for an approximate conservation of mean energies. Unlike 
in the case of stable particles considered in Sec.~\ref{sec:wp-match}, the 
latter does not have a Gaussian form. 
Note the different time dependence of the 
different terms in the exponent in the integrand of the first integral 
in~(\ref{eq:I1}): the terms proportional to $a$ and $b$ are multiplied by 
$t-t_P$, whereas the term $\propto \Gamma$ is multiplied by $t$. This is 
because the former two terms reflect the fact that the peak of the wave packet 
of $P_i$ is located at the neutrino production point $\vec{x}=\vec{x}_P$ at 
the time $t=t_P$, whereas the wave function of this particle exhibits an 
overall exponential suppression starting from its creation time $t=0$. 

Calculating the integral in~(\ref{eq:I1}), we find 
\be
\Phi_{jP}(E_j(\vec{p}),\,\vec{p}) =
\frac{N_P M_{jP}(Q,K)\,}{2\sigma_{pP}^3\, \sigma_{eP}}\,
\exp\!\big[-\frac{(\vec{p}-\vec{P})^2}{4\sigma_{pP}^2} \big]
\exp\!\big[-\frac{\Gamma}{2}t_P-\frac{\tilde{b}^2}{4a}\big]
\Big[\mbox{erf}\Big(\frac{i\tilde{b}}
{2\sqrt{a}}+\sqrt{a}\,t_P\Big)+1\Big]\,,
\label{eq:genwp3} 
\ee
where erf$(x)$ is the error function. The limiting cases of interest can now 
be obtained from the relevant expansions of this function, but it is actually 
easier to study them starting directly with the expression for $I_1$ in 
eq.~(\ref{eq:I1}). 

Indeed, the integrand of $I_1$ contains the exponential and 
oscillating factors, and therefore the integration domain that gives 
significant contribution to $I_1$ is 
\be
|t|\lesssim \min\Big\{
\frac{1}{\sqrt{a}},\,\frac{1}{|\tilde{b}|}\Big\}\,,
\label{eq:domain}
\ee
provided that the right hand side of this condition does not exceed $t_P$;
if it does, for negative $t$ the domain is limited by $|t|>t_P$. 
Let us note that, while the parameter $b$ may vanish, $\tilde{b}$ cannot, as 
it has a non-zero imaginary part $\Gamma/2$. Consider first the limit 
$\sqrt{a}\gg |\tilde{b}|$, $\sqrt{a}\,t_P\gg 1$, which implies  
\be
\sigma_{eP}\gg \Gamma/2\,,\qquad 
\sigma_{eP}\, t_P \gg 1\,.
\label{eq:lim1}
\ee
In this case one can set $\tilde{b}\simeq b$ and also 
extend the lower integration limit in eq.~(\ref{eq:I1}) to $-\infty$, which 
gives $I_1\simeq 
\sqrt{\pi/a}\,\exp[-b^2/4a]$. Substituting this into eq.~(\ref{eq:genwp2}) 
yields, up  to the extra factor $e^{-(\Gamma/2)t_P}$, the old result of 
eqs.~(\ref{eq:genwp}) and ~(\ref{eq:g1}). If instead of the second 
condition in~(\ref{eq:lim1}) one considers the opposite limit $\sigma_{eP}\, 
t_P \ll 1$, the lower integration limit in the last integral 
in~(\ref{eq:I1}) can be set equal to zero.
Since the error function goes to zero for small arguments, if follows that
the result in this 
case is just 1/2 of that in the case $\sigma_{eP}\, t_P \gg 1$. Thus, we 
conclude that in the limit $\sigma_{eP}\gg \Gamma/2$ the approximate 
conservation of mean energies is given by the same (in this case Gaussian) 
law as in the case of the stable neutrino source, with the same energy 
uncertainty $\sigma_{eP}$. This is an expected result.   

Consider now the limit $\sqrt{a}\ll |\tilde{b}|$, $\sqrt{a}\,t_P\ll 1$, or 
\footnote{Note that the second condition in (\ref{eq:lim2}) cannot be relaxed: 
indeed, if one had $\sigma_{eP}\, t_P \gtrsim  1$, then $\Gamma/2\gg 
\sigma_{eP}$ would imply $\Gamma\, t_P \gg  1$, and the wave function of the 
parent particle would be exponentially suppressed by the time the neutrino 
is produced.} 
\be
\sigma_{eP}\ll \Gamma/2\,,\qquad \sigma_{eP}\, t_P \ll 1\,.
\label{eq:lim2}
\ee
In this case one can neglect the term $-a t^2$ in the exponent in the last 
integral in~(\ref{eq:I1}), which yields
\be
I_1\simeq 
e^{-\frac{\Gamma}{2}t_P} 
\frac{i e^{-i\tilde{b} t_P}}{\tilde{b}}
=\frac{i e^{-i b t_P}}{E_j(\vec{p})-E_P-
\vec{v}_P(\vec{p}-\vec{P})+i\Gamma/2}\,.
\label{eq:I1a}
\ee
This is the usual Lorentzian energy distribution factor corresponding to the 
decay of an unstable parent state. 

Thus, we conclude that in the case when $\Gamma/2\ll \sigma_{eP}$ the 
factor in $\Phi_{jP}(E_j(\vec{p}),\vec{p})$ 
that is responsible for 
the approximate conservation of mean energies in the production process  
is essentially the same as in the case of a stable neutrino source 
(Gaussian in the case we considered), whereas in the opposite limit, 
$\Gamma/2\gg \sigma_{eP}$, it is given by the Lorentzian energy distribution 
corresponding to the natural linewidth of the source $\Gamma/2$. In the 
intermediate case $\Gamma\sim \sigma_{eP}$, the energy-dependent factor in 
$\Phi_{jP}(E_j(\vec{p}),\vec{p})$ is neither Gaussian nor Lorentzian, with 
the effective energy uncertainty being of the same order as $\sigma_{eP}$ and 
$\Gamma$.

%=======================================================================
\section{Oscillation probabilities and the normalization conditions 
\label{sec:norm}}
%=======================================================================

%-----------------------------------------------------------------------
\subsection{Normalization of oscillation probabilities in the QM approach
\label{sec:normQM}}
%-----------------------------------------------------------------------

In the QM wave packet approach to neutrino oscillations, one has to 
normalize the oscillation probability $P_{\alpha\beta}(L)$ by hand by 
requiring that it satisfy the unitarity constraint 
\be
{\sum}_\beta P_{\alpha\beta}(L)=1\,.
\label{eq:unitar1}
\ee
This is an ad hoc procedure that is not properly justified within the QM 
formalism; however, it seems to be unavoidable in that approach. In 
particular, one can readily make sure that the standard normalization of the 
neutrino wave packets $\langle\nu_j|\nu_j\rangle=1$, or, equivalently, 
$\int |f_{j}(\vec{p})|^2 d^3p/(2\pi)^3=1$, does not lead to the correct 
normalization of the oscillation probability (that this is indeed the case can 
be easily verified by using Gaussian wave packets as an example). Moreover, 
as we will show below, {\em no independent normalization of the produced and 
detected neutrino states can lead to the correct normalization of the 
oscillation probability in the QM wave packet approach.}

Let us now consider the unitarity constraints on the oscillation probabilities 
and their connection to the normalization conditions in more detail. As it is 
presented in eq.~(\ref{eq:unitar1}), the unitarity condition simply reflects 
the fact that during the propagation between the source and detector neutrinos 
are neither destroyed nor (re)created: once a neutrino is produced, it can 
only change its flavour.%
\footnote{We neglect tiny probabilities of neutrino decay and absorption.} 
Therefore, for a fixed initial flavour $\alpha$, the probabilities 
$P_{\alpha\beta}(L)$ of neutrino conversion to all final flavours $\beta$ sum 
to unity. 
Note that the quantities $P_{\alpha\beta}(L)$ by construction 
depend only on the distance $L$ between the source and the detector and not on 
the propagation time $T$, which is integrated over (see eq.~(\ref{eq:P1})).%
\footnote{For simplicity we assume here the neutrino emission to be spherically 
symmetric. Otherwise, $P_{\alpha\beta}(L)$ has to be defined as 
$P_{\alpha\beta}(L)\equiv\int\! P_{\alpha\beta}(\vec{L}) d\Omega/4\pi$. 
}  
Thus, the meaning of the unitarity condition (\ref{eq:unitar1}) is that, once 
a neutrino is produced, the probability that it will 
be found at a given distance $L$ from the source at some time between zero 
and infinity is equal to one, provided that all flavour states are 
accounted for.  
Similarly, if one introduces $P_{\alpha\beta}(T) \equiv \int d^3L\, 
P_{\alpha\beta}(T,\vec{L})$, it will also satisfy a unitarity condition 
analogous to (\ref{eq:unitar1}). In other words, once the neutrino is created, 
the probability to find it (in any flavour state) at a fixed time $T$ after 
its production somewhere in space 
is equal to one. 

The unitarity conditions can only be satisfied if the oscillation 
probabilities are properly normalized. It is important to note, however, that 
in a consistent formalism unitarity must be satisfied automatically rather 
than being imposed by hand. 

How about the un-integrated probability $P_{\alpha\beta}(T,\vec{L})$, 
should it satisfy a unitarity constraint similar to (\ref{eq:unitar1})? 
Obviously, for arbitrary $\vec{L}$ and $T$ the answer to this question is 
negative. Indeed, 
unless $|\vec{L}-\vec{v}T|\lesssim \sigma_x$ where $\vec{v}$ is the average 
group velocity of the neutrino wave packet and $\sigma_x$ is its spatial 
length, the probabilities $P_{\alpha\beta}(T,\vec{L})$ are vanishingly small 
for all $\alpha$ and $\beta$. Thus, unitarity cannot be used to normalize the 
un-integrated probability. The normalization of $P_{\alpha\beta}(T,\vec{L})$ 
can, however, be fixed differently: in the limit $T\to 0$, $\vec{L}\to 
0$, i.e.\ when the produced neutrino did not have time to evolve yet, the 
probability must satisfy the initial condition $P_{\alpha\beta}(0,\vec{0})=
\delta_{\alpha\beta}$. Note that the above limit should be understood in the 
sense that $\vec{L}$ and $T$ are small compared to the oscillation length but 
large compared to, respectively, the sizes of the spatial localization regions  
and time scales of the neutrino emission and absorption processes. From 
eqs.~(\ref{eq:P0}) and (\ref{eq:Aj}) it then immediately follows that the 
amplitudes ${\cal A}_j(T,\vec{L})$ corresponding to neutrino mass eigenstates 
must satisfy ${\cal A}_j(0,\vec{0})=e^{i\phi}$ where the real phase $\phi$ 
is the same for all $j$. By a rephasing of the momentum distribution functions 
of either emitted or detected neutrinos, this common phase can  be eliminated,  
and one finally gets ${\cal A}_j(0,\vec{0})=1$. 

Now, let us check if this condition is fulfilled in the particular case of 
Gaussian wave packets. For the momentum distribution functions of the produced 
and detected neutrino states normalized according to eq.~(\ref{eq:norm2}), 
i.e.\ having the form~(\ref{eq:Gauss1}), a straightforward calculation yields
\be
{\cal A}_j(0,\vec{0})=\int\!\frac{d^3p}{(2\pi)^3}\, f_{jP}(\vec{p},\vec{P}) 
\, f^*_{jD}(\vec{p},\vec{P}') =\left(\frac{2\sigma_{pP}\sigma_{pD}}
{\sigma_{pP}^2+\sigma_{pD}^2}\right)^{3/2} 
\exp\Big[-{\frac{(\vec{P}-\vec{P}')^2}{4(\sigma_{pP}^2+\sigma_{pD}^2)}}
\Big]\,. 
\label{eq:overlap1}
\ee
For 
$\sigma_{pP}\ne \sigma_{pD}$ and $\vec{P}\ne \vec{P}'$ both factors on the 
right hand side of the last equality are smaller than one, and therefore the 
condition ${\cal A}_j(0,\vec{0})=1$ is clearly violated. Moreover, the 
dependence of the result on the parameters of the produced and detected 
neutrino states does not factorize; this means that no independent 
normalization of these states can lead to the correct normalization of the 
amplitude, as pointed out above.   

It is actually quite easy to understand why this happens. The integral in 
eq.~(\ref{eq:overlap1}) is nothing but the overlap integral of the wave 
functions of the produced and detected neutrinos. If these wave functions are 
normalized to unity and the momentum (or energy) spectra of the emitted and 
detected states do not coincide, this overlap integral is always less than 
one. In reality, the spectra of the emitted and absorbed neutrino states are 
determined by the physical nature and experimental conditions of the neutrino 
production and detection processes, which are always different. This, in 
particular, means that a fraction of the produced neutrinos may simply not be 
detectable. For instance, if the threshold in the detection process (either 
the physical threshold or the one imposed by energy cuts of the detected 
events) is higher than the maximum energy of the emitted neutrino, no 
detection will be possible at all. Mathematically, the fact that the overlap 
integral~(\ref{eq:overlap1}) is always less than one is a consequence of the 
Schwarz inequality $|(f,g)|^2\le (f,f)(g,g)$, where the equality is only 
reached if $f=const\cdot g$. 

Even if one adopts the unrealistic assumption $f_{jP}(\vec{p})=f_{jD}(\vec{p})$ 
(which for Gaussian wave packets would mean $\vec{P}=\vec{P}'$ and 
$\sigma_{pP}=\sigma_{pD}$), this will not solve all the normalization problems 
of the QM wave packet approach. The condition ${\cal A}_j(0,\vec{0})=1$ will 
be satisfied in this case; however, the physically observable oscillation 
probability $P_{\alpha\beta}(\vec{L})$ defined in eq.~(\ref{eq:P1}) will 
still not be properly normalized, and the unitarity condition 
(\ref{eq:unitar1}) will not be satisfied. Indeed, from eq.~(\ref{eq:P0}) 
it follows that unitarity requires that \cite{AS} 
\be
\int dT \,|{\cal A}_j(T, \vec{L})|^2 =1\,
\label{eq:unitar2}
\ee
for all $j$. Obviously, the fulfilment of the condition ${\cal A}_j(0,\vec{0})
=1$ does not enforce (\ref{eq:unitar2}); therefore in the QM 
formalism condition (\ref{eq:unitar2}) has to be imposed by hand. 

It is not difficult to understand why yet another normalization problem 
arises: integration over the time $T$ should actually be considered as time  
averaging in the QM approach, and the integral on the right hand side 
of eq.~(\ref{eq:P1}) should be normalized by dividing it by the characteristic 
time $\Delta T$ which depends on time scales of both the neutrino production 
and detection processes. This follows from the fact that the amplitude 
${\cal A}_j(T, \vec{L})$ is substantially different from zero only when 
$|\vec{L}-\vec{v}T|\lesssim \sigma_x$, where the effective length of the 
wave packet $\sigma_x$ is determined by both the neutrino production and 
detection processes, which gives $\Delta T\sim \sigma_x/v$. 
It is difficult to calculate the quantity $\Delta T$ precisely, and 
the simplest way out is just to impose the unitarity condition by hand, 
which yields the correct normalization of $P_{\alpha\beta}(L)$.%
\footnote{Note that once the normalization condition (\ref{eq:unitar2})
is enforced, one can demonstrate that the resulting oscillation probability
$P_{\alpha\beta}(L)$ is Lorentz invariant \cite{AS}. This is not trivial in 
the QM approach because the QM formalism is not manifestly Lorentz covariant.}

Thus, in the QM approach there are two sources of the normalization problems: 
lack of the overlap of the wave functions of the produced and 
detected neutrino states and the necessity of integration over the 
neutrino propagation time $T$. We will show now how both these problems 
are naturally solved in the QFT-based formalism.

%-----------------------------------------------------------------------
\subsection{Normalized oscillation probabilities in the QFT framework
\label{sec:normQFT}}
%-----------------------------------------------------------------------

\subsubsection{Generalities \label{sec:gener}}
%---------------------------------------------

Let us start with recalling the operational definition of the neutrino 
oscillation probability. In a detection process that is sensitive to neutrinos 
of flavour $\beta$, the detection rate is 
\footnote{We omit the obvious factors of detection efficiency and energy 
resolution that are not relevant to our argument.} 
\be
\Gamma^{\rm det}_\beta=\int dE \,j_\beta(E)\sigma_\beta(E)\,,
\label{eq:GammaD}
\ee
where $\sigma_\beta(E)$ is the detection cross section and $j_\beta(E)$ 
is the energy density (spectrum) of the $\nu_\beta$ flux at the detector. 
If a source at a distance $L$ from the detector emits neutrinos of 
flavour $\alpha$ with the energy spectrum $d\Gamma_\alpha^{\rm prod}(E)/dE$, 
the energy density of the $\nu_\beta$ flux at the detector is 
\be
j_\beta(E)=\frac{1}{4\pi L^2}\frac{d\Gamma_\alpha^{\rm prod}(E)}{dE}
P_{\alpha\beta}(L,E)\,,
\label{eq:jbeta}
\ee 
where $P_{\alpha\beta}(L,E)$ is the $\nu_a\to\nu_b$ oscillation 
probability for neutrinos of energy $E$, and we once again assumed for 
simplicity neutrino emission to be spherically symmetric. Substituting 
this into eq.~(\ref{eq:GammaD}) yields the rate of the overall neutrino 
production-propagation-detection process: 
\be
\Gamma^{\rm tot}_{\alpha\beta}\,\equiv\int dE\, 
\frac{d\Gamma^{\rm tot}_{\alpha\beta}(E)}{dE}\,=\,\frac{1}{4\pi L^2}
\int dE\,\frac{d\Gamma_\alpha^{\rm prod}(E)}{dE}\,P_{\alpha\beta}(L,E)\,
\sigma_\beta(E)\,. 
\label{eq:GammaTot}
\ee
The oscillation probability can now be extracted from the integrand of 
eq.~(\ref{eq:GammaTot}) by dividing it by the neutrino emission 
spectrum, detection cross section and the geometrical factor $1/4\pi L^2$:
\be
P_{\alpha\beta}(L,E)\,=\,
\frac{d\Gamma^{\rm tot}_{\alpha\beta}(E)/dE}{\frac{1}{4\pi L^2}\,
[d\Gamma_\alpha^{\rm prod}(E)/dE]\,\sigma_\beta(E)}\,. 
\label{eq:P2}
\ee
Note that an important ingredient of this argument is the assumption that 
at a fixed neutrino energy $E$ the overall rate of the process factorizes 
into the production rate, propagation (oscillation) probability and detection 
cross section. Should such a factorization turn out to be impossible, the 
very notion of the oscillation probability would lose its sense, and one 
would have to deal instead with the overall rate of neutrino production, 
propagation and detection. 

Now let us come back to the QFT-based treatment of neutrino oscillations 
and try to cast the rate of the overall process in the form of 
eq.~(\ref{eq:GammaTot}). To this end, we return to eq.~(\ref{eq:amp6}) 
for the amplitude of the process but, unlike in the previous sections, 
perform in it the integration over 3-momentum before integrating over the 
energy variable $p^0\equiv E$. In doing so, we will make use of a result 
obtained by Grimus and Stockinger \cite{Grimus:1996av}, which states 
that, for a large baseline $L$, positive $A$ and a sufficiently smooth 
function $\psi(\vec{p})$, 
\be
  \int d^3p \frac{\psi(\vec{p}) \, e^{i \vec{p} \vec{L}}}{A - \vec{p}^2 
+ i\epsilon}
    = -\frac{2 \pi^2}{L} \psi(\sqrt{A} \tfrac{\vec{L}}{L}) e^{i \sqrt{A}L}
      + \mathcal{O} (L^{-\frac{3}{2}})\,,
\label{eq:Grimus}
\ee
whereas for $A<0$ the integral behaves as $L^{-2}$. This result was obtained 
in \cite{Grimus:1996av} in the limit $L\to \infty$, but a careful 
examination of the derivation shows that its applicability condition is 
actually $L\gg p_j/\sigma_p^2$, which we will assume to be satisfied. 
Applying (\ref{eq:Grimus}) to eq.~(\ref{eq:amp6}) yields   
\be
i{\cal A}_{\alpha\beta}(T,\vec{L})=\frac{-i}{8\pi^2 L} \sum_j U_{\alpha 
j}^* U_{\beta j}^{} \int\!dE\,\Phi_{jP}(E, p_j\vec{l})\Phi_{jD}(E, p_j\vec{l})\,
\,2 E\,e^{-i E\, T + i p_j L}\,,
\label{eq:amp}
\ee
where 
\be
p_j\equiv\sqrt{E^2-m_j^2}\,,\qquad\quad \vec{l}\equiv\frac{\vec{L}}{L}\,.
\label{eq:not2}
\ee
Next, we note that, 
just as the quantities $\Phi_{jP}(E_j,\vec{p})$ depend on the index $j$ only 
through the neutrino energy $E_j$, the functions $\Phi_{jP}(E,p_j\vec{l})$ 
depend on the index $j$ only through the neutrino momentum $p_j$. 
Therefore, to simplify the notation we will denote
\be
\Phi_{jP}(E,p_j\vec{l})\equiv \Phi_{P}(E,p_j\vec{l})\,,
\qquad\quad 
\Phi_{jD}(E,p_j\vec{l})\equiv \Phi_{D}(E,p_j\vec{l})\,.
\label{eq:not3}
\ee 
The overall probability of the neutrino production-propagation-detection 
process $P_{\alpha\beta}^{\rm tot}(T,\vec{L})$ is the squared modulus of 
the amplitude~(\ref{eq:amp}):
\be
P_{\alpha\beta}^{\rm tot}(T, \vec{L})\equiv 
|{\cal A}_{\alpha\beta}(T,\vec{L})|^2=
\sum_{j,k} U_{\alpha j}^* U_{\beta j}^{} 
U_{\alpha k}^{} U_{\beta k}^* \, {\cal A}_j(T,\vec{L}) \,
{\cal A}_k^*(T,\vec{L})\,.
\label{eq:PTL}
\ee 	
We will actually need the integral of this probability over the time $T$ 
(the reasons for integration over $T$ will be discussed in 
Secs.~\ref{sec:steady} and~\ref{sec:disc}):
\begin{align}
\tilde{P}^{\rm tot}_{\alpha\beta}(\vec{L})
=\int dT\,P_{\alpha\beta}(T,\vec{L})=&
\frac{1}{8\pi^2}\frac{1}{4\pi L^2}
\sum_{j,k}U_{\alpha j}^* U_{\beta j}^{}
U_{\alpha k}^{} U_{\beta k}^*
\nonumber \\
\times \int\!dE\, \Phi_{P}(E, p_j\vec{l}) &\Phi_{D}(E, p_j\vec{l})\,
\Phi^*_{P}(E, p_k\vec{l}) \Phi^*_{D}(E, p_k\vec{l})\,
\,(2 E)^2\,e^{i (p_j-p_k) L}\,.
\label{eq:Ptot1}
\end{align}
We use the tilde here to stress that $\tilde{P}_{\alpha\beta}^{\rm tot}
(\vec{L})$ is not a probability but rather a time-integrated probability, 
which has the dimension of time. 
Note that (\ref{eq:Ptot1}) contains an incoherent sum (integral) over 
contributions from different energy eigenstates. This means that only the 
amplitudes corresponding to the same neutrino energy interfere, as it 
is in the stationary case. This is related to the integration over time $T$ 
and is a reflection of the fact that the time-integrated nonstationary 
probability is equivalent to the energy-integrated stationary probability 
\cite{Beuthe1,AS}. 
It should also be stressed that, although the integration over $E$ in 
eq.~(\ref{eq:Ptot1}) is formally performed over the interval 
$(-\infty,\, \infty)$ (recall that $E$ coincides with the variable $p^0$ 
of eq.~(\ref{eq:amp6})), the contribution of the unphysical region of 
negative energies is actually negligible and can be discarded. This is a 
consequences of the fact that $\Phi_{P,D}$ are sharply peaked at  
positive values of energy $E_{P,D}$, with the peak widths satisfying 
$\sigma_{eP,eD}\ll E_{P,D}$. 

To simplify the following consideration, we will once again assume that 
the neutrino production and detection processes are isotropic. In our 
approach this means that we have to average the quantities 
$\Phi_{P}(E,p_j\vec{l})$ and $\Phi_{D}(E,p_j\vec{l})$ over the 
directions of the incoming particles $P_i$ and $D_i$, which amounts to 
averaging over the directions of $\vec{l}$. We can therefore denote 
$\Phi_{P,D}(E,p_j)\equiv \int\frac{d\Omega_{\vec{l}}}{4\pi}
\Phi_{P,D}(E,p_j\vec{l})$ and drop $\vec{l}$ from the arguments of 
$\Phi_{P,D}$ in eq.~(\ref{eq:Ptot1}) and all the subsequent expressions. 
Relaxing the isotropy assumption would complicate the analysis but would 
not change the final result for the probability of neutrino 
oscillations.    
  
The next step is to calculate the neutrino production and detection 
probabilities.  As can be seen  from eq.~(\ref{eq:amp6}), neutrinos in the 
intermediate state are considered in our framework as plane waves weighted 
with the factors $\Phi_{P,D}$. We therefore describe the external particles 
by wave packets and neutrinos by plane waves. Application of the standard 
rules of QFT then yields 
\footnote{Note that we do not include the factor $[2E_j(p)]^{-1}$ in 
the integration measure because it is already included in the 
definition of $|\Phi_{jP}(E,\vec{p})|^2$, see eqs.~(\ref{eq:M}) and 
(\ref{eq:psi}).} 
\begin{align}
P^{\rm prod}_\alpha =\sum_j |U_{\alpha j}|^2 \int &
\frac{d^3 p_j}{(2\pi)^3}\,
\big|\Phi_{P}(E, p_j)\big|^2 
\nonumber \\ &
=\sum_j |U_{\alpha j}|^2 \frac{1}{8\pi^2}\int dE 
\,\big|\Phi_{P}(E,p_j)\big|^2 4 E p_j\,. 
\label{eq:prodrate}
\end{align}
The spectral density of emitted neutrino flux, $dP_\alpha^{\rm prod}(E)/dE$, 
is obtained by removing the integration over energy on the right hand side of 
the last equality in (\ref{eq:prodrate}). 

For the detection probability we obtain 
\be
P_\beta^{\rm det}(E)=\sum_k |U_{\beta k}|^2 |\Phi_{D}(E, p_k)|^2 \frac{1}
{V}\,,
\label{eq:detProb}
\ee
where the normalization volume $V$ comes from the plane-wave description of 
the incoming neutrino. Note that the expression for the production probability 
$P_\alpha^{\rm prod}$ in eq.~(\ref{eq:prodrate}) does not contain the 
factor $1/V$ even though it is also calculated for plane-wave neutrinos. 
This is because neutrinos are in the final state at production, and the 
calculation of their phase space volume involves integration over $V 
d^3 p$.

\subsubsection{The case of continuous fluxes of incoming particles 
\label{sec:steady}}
%-----------------------------------------------------------------

\noindent
A direct inspection of the expressions for the probabilities of neutrino 
production and detection as well as of the probability of the overall 
production-propagation-detection process 
obtained above shows that they are independent of the total running time 
of the experiment $t$. This is because they were calculated for 
individual processes with single external wave packets of each type, and 
the microscopic production and detection time intervals were assumed to be 
centered at fixed instants of time $t_P$ and $t_D$, respectively. 
On the other hand, 
in practice one is usually interested in the total probabilities for the 
processes to occur within a macroscopic time interval of length $t$ or in 
interaction rates. Normally, the probabilities are proportional to $t$, 
while the rates are $t$-independent. In the wave packet approach, this can be 
achieved if we take into account that in realistic situations one often 
has to deal with continuous fluxes of incoming particles. (We will comment on 
the opposite case of stationary initial states in the next subsection.)

Let us start with calculating the production rate and detection cross section 
in the case of steady fluxes of the incoming particles. Consider some interval 
of time $T_0$ that is large compared to the time scales of the neutrino 
production and detection processes. Let the number of the projectile 
particles $P_i$ entering the production region (e.g.\ a spherical region of 
radius $\sigma_{xP}$ around the point $\vec{x}=\vec{x}_P$) during this 
interval be $N_P$. Then the number of particles $P_i$ entering the 
production region during the interval $d t_P$ is $d N_{P_i} = N_P (dt_P/T_0)$. 

If the production probability in the case of the individual process with 
single external wave packets is $P_\alpha^{\rm prod}$, the probability of 
neutrino emission during the finite interval of time $t$ ($0\le t\le T_0$) 
is 
\footnote{\label{foot:prod} 
If the flux of the incoming particles is not steady, the number 
of particles entering the production region over the time $t$ is given by 
$\int_0^t \rho_P(t_P) dt_P$, where $\rho_P(t_P)$ is the distribution of 
these particles with respect to $t_P$, normalized according to $\int_0^{T_0} 
\rho_P(t_P) dt_P=N_P$. The right hand side of the first equality in 
eq.~(\ref{eq:Pprod}) then has to be replaced by $\int_0^t \rho_P(t_P) 
P_\alpha^{\rm prod} dt_P$. 
For a steady flux the distribution is uniform, i.e.\ $\rho_P(t_P)=N_P/T_0=
const$.} 
\be
{\cal{P}}_\alpha^{\rm prod}(t)=N_P\int_0^t \frac{dt_P}{T_0}\,P_\alpha^{\rm prod}
=N_P \,P_\alpha^{\rm prod} \frac{t}{T_0}\,.
\label{eq:Pprod}
\ee 
The integration is trivial because $P_\alpha^{\rm prod}$ is actually 
independent of $t_P$ due to invariance with respect to time translations. 
Note that the probability ${\cal P}_\alpha^{\rm prod}$ is proportional to 
$t$. We can therefore define the production rate in the usual way:
\be
\Gamma_\alpha^{\rm prod}=\frac{d{\cal P}_\alpha^{\rm prod}(t)}{dt}=
N_P \,\frac{P_\alpha^{\rm prod}}{T_0}\,.
\label{eq:Prodrate2}
\ee 

Let us now consider the detection cross section. Normally, a cross 
section is defined for a single, fixed target particle.
However, the initial-state particles $D_i$ in our treatment of the detection 
process are described by moving wave packets, which enter the detection region
that is centered at a fixed point $\vec{x}=\vec{x}_D$. Therefore, 
our treatment of neutrino detection should be similar to that  of the 
production process. 
For the individual detection process with single external wave packets 
of each type the detection probability is given by eq.~(\ref{eq:detProb}).   
Let us now assume that the number of the particles $D_i$ 
entering the detection region during the interval of time $T_0$ is $N_D$.  
Then we obtain for the time-dependent detection probability in the case of 
steady incoming fluxes of $D_i$ and neutrinos 
\be
{\cal{P}}_\beta^{\rm det}(t)=N_D\int_0^t\frac{dt_D}{T_0}\,P_\beta^{\rm 
det}=N_D \,P_\beta^{\rm det} \frac{t}{T_0}\,,
\label{eq:detProb2}
\ee 
and for the detection rate 
\be
\Gamma_\beta^{\rm det}=\frac{d{\cal P}_\beta^{\rm det}(t)}{dt}=
N_D \,\frac{P_\beta^{\rm det}}{T_0}\,.
\label{eq:detrate2}
\ee 
To obtain the detection cross section we have to divide this rate (more 
precisely, the summand of the sum over $k$ that enters into 
(\ref{eq:detrate2})) by the flux of incoming neutrinos $j_{\nu k}=n_{\nu k} 
v_{\nu k}$, where $n_{\nu k}$ is the number density of the detected $\nu_k$ 
and $v_{\nu k}$ is their velocity. With our normalization (one particle in
the normalization volume) we have $n_{\nu k}=1/V$, and from
eqs.~(\ref{eq:detProb}) and (\ref{eq:detrate2}) we finally obtain 
\be
\sigma_\beta(E)= \frac{N_D}{T_0} \sum_k |U_{\beta k}|^2 
|\Phi_{D}(E,p_k)|^2 \frac{E}{p_k}\,.
\label{eq:sigmabeta}
\ee

Now we proceed to the calculation of the rate of the overall 
production-propagation-detection process. Since we want to calculate this 
quantity not for a single process with individual wave packets of external 
particles but for steady fluxes of incoming $P_i$ and $D_i$, we have to 
integrate the $T$-dependent probability of the single process 
$P_{\alpha\beta}^{\rm tot}(T,L)$ given by eq.~(\ref{eq:PTL})
over both $t_P$ and $t_D$. Proceeding in the same way as before, we find 
for the probability of the process for steady fluxes of the incoming 
particles  
\be
{\cal P}_{\alpha\beta}^{\rm tot}(t,L)=\frac{N_P N_D}{T_0^2}\int_0^t dt_D 
\int_0^t dt_P \, P_{\alpha\beta}^{\rm tot}(T,L)\,.
\label{eq:probtot1}
\ee
Introducing the new integration variables $\tilde{T}\equiv (t_P+t_D)/2$ 
and $T=t_D-t_P$, we obtain 
\begin{align}
{\cal P}_{\alpha\beta}^{\rm tot}(t,L)=&\,
\frac{N_P N_D}{T_0^2}\,\Big[
\int_0^t dT \,P_{\alpha\beta}^{\rm tot}(T,L)(t-T)+ 
\int_{-t}^0 dT \,P_{\alpha\beta}^{\rm tot}(T,L) (t+T)\Big]
\nonumber \\
=&\,\frac{N_P N_D}{T_0^2}\,\Big[
t \int_{-t}^t dT \,P_{\alpha\beta}^{\rm tot}(T,L)-
\int_0^t dT \, T P_{\alpha\beta}^{\rm tot}(T,L)
+\int_{-t}^0 dT \, T P_{\alpha\beta}^{\rm tot}(T,L) \Big]
\nonumber \\
\equiv & \,\frac{N_P N_D}{T_0^2}\,\Big[
t I_1(t) - I_2(t) + I_3(t) \Big] \,. 
\label{eq:probtot2}
\end{align}
It can be readily shown that in the limit of large $t$ (much larger than the 
time scales of the neutrino production and detection processes) the integral 
$I_1$ coincides with the quantity $\tilde{P}_{\alpha\beta}^{\rm tot}(L)$ 
defined in eq.~(\ref{eq:Ptot1}), whereas $I_2$ and $I_3$ give negligible 
contributions (see Appendix~B). Therefore for large $t$ 
eq.~(\ref{eq:probtot2}) can be rewritten as 
\be
{\cal P}_{\alpha\beta}^{\rm tot}(t,L)=
\frac{N_P N_D}{T_0^2}\,t\,\tilde{P}_{\alpha\beta}^{\rm tot}(L)\,.
\label{eq:probtot3}
\ee
The rate of the overall process is then 
\footnote{
For the reader willing to check the dimensions of our 
expressions, we note that from the definitions of $\Phi_{P,D}$ it follows 
that they have dimension $m^{-3/2}$. It is then easy to see that the 
amplitude (\ref{eq:amp}) and the probabilities 
(\ref{eq:Ptot1})-(\ref{eq:Pprod}), 
(\ref{eq:detProb2}), (\ref{eq:probtot1}) and (\ref{eq:probtot3}) are 
dimensionless, the cross section (\ref{eq:sigmabeta}) has the dimension 
of squared length (or $m^{-2}$), and the rates  (\ref{eq:Prodrate2}), 
(\ref{eq:detrate2}) and (\ref{eq:totrate}) have the dimension of 
inverse time (or $m$), as they should.}
\be
\Gamma_{\alpha\beta}^{\rm tot}(L)=\frac{d{\cal P}_{\alpha\beta}^{\rm 
tot}(t,L)}{dt}=N_P N_D \,\frac{\tilde{P}_{\alpha\beta}^{\rm tot}}{T_0^2}\,.
\label{eq:totrate}
\ee

\subsubsection{The case of stationary initial states \label{sec:stationary}}
%---------------------------------------------------------------------------

If the initial state particles $P_i$ and/or $D_i$ are in stationary states 
rather than being described by moving wave packets, the above consideration 
has to be slightly modified. Consider, for example, neutrino production 
in decays of unstable particles $P_i$ bound in a solid. 
Let $\rho_P(t_P)$ be the probability distribution function for the decay times
of the parent particles in the source, and let $N_P$ be defined by the
condition $\int_0^{T_0} \rho_P(t_P) dt_P=N_P$. If $T_0$ is short compared to
the lifetime $\Gamma^{-1}$ of $P_i$, one has $\rho_P(t_P)=const=N_P/T_0$.
If, on the contrary, $T_0 \gtrsim \Gamma^{-1}$, $\rho_P(t_P)$ will usually have
an exponential form.%
\footnote{An exception is the case where $P_i$ in the source are continuously
replenished. In this case, the function $\rho_P(t_P)$ will depend on the time
dynamics of the production of these particles.} 
The neutrino production probability is then again given by 
eq.~(\ref{eq:Pprod}), just as in the case of a continuous flux of incoming 
particles $P_i$.  

The situation with bound-state stationary particles $D_i$ in the detector
can be considered quite similarly. One can assume $D_i$ to be stable. If 
the source creates a steady flux of neutrinos, then for the 
ensemble of $D_i$ in the detector the distribution $\rho_D(t_D)$ of the 
detection times $t_D$ is uniform and is given by $N_D/T_0$, where $N_D$ is 
defined by the normalization condition $\int_0^{T_0} \rho_D(t_D) dt_D=N_D$. 
The detection probability is then given by eq.~(\ref{eq:detProb2}). 

Thus, with these re-interpretations of $N_P$ and $N_D$, the expressions 
for the neutrino production and detection probabilities and rates and 
the detection cross section obtained in the previous subsection remain valid 
in the case of stationary initial states as well. 

We are now in a position to obtain the normalized oscillation 
probability.

\subsubsection{The oscillation probability in the QFT approach 
\label{sec:normprob}}
%-------------------------------------------------------------

In the case when the rate of the overall production-propagation-detection 
process at a fixed neutrino energy factorizes, the oscillation probability 
should be obtainable from eq.~(\ref{eq:P2}).
Substituting eqs.~(\ref{eq:totrate}) and (\ref{eq:Prodrate2}) 
(with $\tilde{P}_{\alpha\beta}^{\rm tot}$ and $P_{\alpha}^{\rm prod}$ defined 
in eqs.~(\ref{eq:Ptot1}) and (\ref{eq:prodrate})) and eq.~(\ref{eq:sigmabeta}) 
into this relation, we find 
\be
``P_{\alpha\beta}(L,E)"\,=\,\frac{\sum_{j,k}U_{\alpha j}^* U_{\beta j}^{} 
U_{\alpha k}^{} U_{\beta k}^* 
\Phi_{P}(E, p_j) \Phi_{D}(E, p_j)\,\Phi^*_{P}(E, p_k) 
\Phi^*_{D}(E, p_k)\,\,e^{i (p_j-p_k) L}}{\sum_j |U_{\alpha j}|^2 \,
|\Phi_{P}(E,p_j)|^2\,p_j\,\sum_k |U_{\beta k}|^2 \,|\Phi_{D}(E,p_k)|^2 
p_k^{-1}} \,. 
\label{eq:P3}
\ee
The quotation marks here are to remind us that we yet have to prove that this 
quantity can indeed be interpreted as the oscillation probability. 

The alert reader has probably noticed that, while the integral in 
(\ref{eq:GammaTot}) is taken over the energies of different neutrinos in 
the neutrino flux, the integration in (\ref{eq:Ptot1}) is performed over 
the energy distribution within the wave packet of an individual neutrino.
This, however, does not invalidate our argument leading to eq.~(\ref{eq:P3}). 
The reason for this is that the following two situations 
are known to be experimentally indistinguishable \cite{Rich:1993wu,Kiers}: 
(a) a flux of neutrinos described by identical wave packets, each with an
energy spread $f(E)$, and (b) a flux of neutrinos, each with a sharp energy, 
with the overall energy distribution $\phi(E)=|f(E)|^2$. 

Let us now discuss the conditions under which eq.~\eqref{eq:P3} is a
well-defined oscillation probability. For this to take the place,  
the expression for the differential rate $d\Gamma^\text{tot}/dE$ of 
the overall process should 
factorize into the production rate, the oscillation probability, and the
detection cross section. This means that it should be possible to pull the
differential flux and the detection cross section out of the numerator of
eq.~\eqref{eq:P3} in analogy to eq.~\eqref{eq:GammaTot}, and then cancel 
them against the denominator.  This, in turn, requires that the momentum 
distributions $\Phi_{P}(E,p_j)$ and $\Phi_{D}(E,p_j)$ be virtually independent 
of the neutrino mass eigenstate index $j$. To see when this is the case, we 
first note that the momentum distribution functions $\Phi_P$ are all peaked at 
the same momentum $P$.  Therefore, if $|p_j-p_k|$ is much smaller than the 
width of the peak $\sigma_{pP}$, we can replace the factors $\Phi_{P}(E,p_j)$ 
in eq.~\eqref{eq:P3} by the common value $\Phi_{P}(E,p)$ calculated at the 
average momentum $p$, pull them out of the sums in the numerator and in 
the denominator, and cancel them.\footnote{It should be stressed that the mean
momentum $p$ is defined here as an average over different mass eigenstates of
the momenta $p_j = (E^2-m_j^2)^{1/2}$ taken at the same fixed value of energy
$E$. It is therefore different from the mean momentum $P$ of the individual
wave packets, introduced earlier, for which the average was taken over the
spread of momenta (or energies) within the wave packet.} A similar argument
applies to the momentum distribution functions associated with the detection
process, $\Phi_D$, which are all peaked at the same momentum $P'$ and have
widths $\sigma_{pD}$. When neutrinos are ultra-relativistic or quasi-degenerate
in mass, i.e.\ when
\begin{align}
  |p_j-p_k| \ll p_j, p_k \,,
  \label{eq:ultrarel}
\end{align}
we can also replace $p_j$, $p_k$ by $p$ in the denominator of
eq.~\eqref{eq:P3}.%
\footnote{Note that condition \eqref{eq:ultrarel} is usually weaker than 
the requirement $|p_j-p_k| \ll \sigma_{pP}, \sigma_{pD}$. The phase
space regions where it is stronger, i.e.\ where $p_j, p_k < \sigma_{pP},
\sigma_{pD}$, are usually far from the peaks of the momentum distribution
functions and are therefore suppressed.} 
We can then use unitarity of the
leptonic mixing matrix to reduce eq.~\eqref{eq:P3} to
\begin{equation}
  P_{\alpha\beta}(L,E)\,=\, {\sum}_{j,k}U_{\alpha j}^* U_{\beta j}^{}
  U_{\alpha k}^{} U_{\beta k}^* \,e^{i (p_j-p_k) L}\,.
  \label{eq:P3a}
\end{equation}
Since for ultra-relativistic or quasi-degenerate neutrinos $p_j-p_k\simeq
-\Delta m_{jk}^2/2p$, this is just the standard formula for the probability of
neutrino oscillations in vacuum. Thus, the QFT-based 
approach allows one to identify the conditions under which 
$P_{\alpha\beta}(L,E)$ can be sensibly defined, and also gives the correctly 
normalized  expression for this probability. The conditions are that
neutrinos should be ultra-relativistic or quasi-degenerate in mass and, in
addition, the inequality
\begin{equation}
  |p_j - p_k| \simeq \frac{\Delta m_{jk}^2}{2p}\ll \sigma_p\,
  \label{eq:cond3a}
\end{equation}
should be satisfied.
Here, we have introduced the effective momentum uncertainty $\sigma_p$,
which is dominated by the smallest between
$\sigma_{pP}$ and $\sigma_{pD}$.  Since $\sigma_{pP}$ and $\sigma_{pD}$, in
turn, are dominated by the energy uncertainties $\sigma_{eP}$ and
$\sigma_{eD}$, repectively (see Sec.~\ref{sec:wp-match}),
condition~\eqref{eq:cond3a} is equivalent to the one in eq.~\eqref{eq:cond2}.

If condition~\eqref{eq:cond3a} is violated, at least one of the momentum
distributions $\Phi_P(E, p_j)$, $\Phi_P(E, p_k)$, $\Phi_D(E, p_j)$, $\Phi_D(E,
p_k)$ will be strongly suppressed. This implies that the numerator of 
eq.~\eqref{eq:P3} will not factorize in accordance with 
eq.~\eqref{eq:GammaTot} in this case, so that the oscillation probability will 
not be a well-defined quantity. Also, eq.~\eqref{eq:P3} will in general not 
satisfy the unitarity condition~\eqref{eq:unitar1}. It is important that the 
interference terms involving the suppressed momentum distributions in the 
numerator of eq.~\eqref{eq:P3} will be quenched in this case, and thus neutrino 
oscillations involving the corresponding mass eigenstates will be inhibited. 
Physically, this can be traced to the lack of coherence at neutrino production
and/or detection. It can be shown that production or detection decoherence is
equivalent to the lack of localization of, respectively, the production or
detection process \cite{Kayser:1981ye, Rich:1993wu, AS}.\footnote{While
condition (\ref{eq:cond3a}) ensures the production/detection coherence
(localization), it says nothing about another possible source of decoherence --
separation of neutrino wave packets at long enough distances $L>L_{\rm coh}$
due to the difference of the group velocities of different neutrino mass
eigenstates. This is related to the fact that a fixed neutrino energy
corresponds to the stationary situation, when the coherence length $L_{\rm
coh}\to \infty$. The finite coherence length is recovered upon the integration
over energy in eq.~(\ref{eq:Ptot1}) \cite{Beuthe1}.}

The above QFT-based considerations also allow one to shed some light on the
meaning of the normalization condition imposed on the oscillation probability
in the QM wave packet approach, which looks rather arbitrary within the QM
framework. As was pointed out in Sec.~\ref{sec:qft-wp}, the QM and QFT
approaches can be matched if the QM quantities $f_{jP}$ and $f_{jD}$ are
identified with the QFT functions $\Phi_{jP}(E_j,\vec{p})$ and
$\Phi_{jD}^*(E_j,\vec{p})$, respectively. The latter, however, bear information
not only on the properties of the emitted and absorbed neutrinos, but also on
the production and detection processes. The QM normalization procedure that is
tailored to obtain an expression for the oscillation probability satisfying the
unitarity condition can then be easily seen to be equivalent, in the
limit~\eqref{eq:cond3a}, to the division of the overall rate of the process by
the production rate and detection cross section, as in eq.~\eqref{eq:P3}.

%=======================================================================
\section{Some additional comments \label{sec:addit}}
%=======================================================================

In this section we will comment on some issues pertaining to the description 
of neutrino oscillations in the external and intermediate wave packet approaches 
that were not discussed or were only briefly mentioned above. We will 
also show how one can relax some assumptions usually adopted in the QM and QFT 
approaches.

\vspace*{1.2mm}
\noindent
{\em 1. Unequal mean momenta of the produced and detected 
neutrino states}

In the vast majority of derivations of the neutrino oscillation 
probability within the QM wave packet framework, it was assumed that the 
mean momenta of the produced and detected neutrino states, $\vec{P}$ and 
$\vec{P}'$, coincide.  
\footnote{The only exceptions we are aware of are refs.~\cite{AS,Kopp:2009fa}.} 
There are, however, no reasons for this to be the case. Indeed, the mean 
momenta of the emitted and absorbed neutrino states are determined by the 
kinematics and experimental conditions of the neutrino production and 
detection processes, respectively, and those are in general different. 
Let us now examine the consequences of $\vec{P}\ne \vec{P}'$ using, as 
before, the case of Gaussian wave packets as an example. 

It was shown in Sec.~\ref{sec:wp-match} that the neutrino wave packets
derived in QFT can be cast into the form usually adopted for neutrino wave packets 
in the QM approach if one uses for the latter the effective (in general 
anisotropic) momentum uncertainties, shifted mean momenta and modified 
normalization factors. Since the expressions for the QM wave packets look 
simpler, we will study the implications of  unequal mean momenta of 
the produced and detected neutrino states within the QM formalism. 
We will also ignore possible anisotropy of the neutrino 
momentum uncertainties. Taking it into account would just complicate the 
calculations without changing the essence of the result.

We will now calculate the amplitude (\ref{eq:Aj}) corresponding to the 
emission and absorption of the neutrino mass eigenstate $\nu_j$. To do 
so, we expand the energy $E_j(\vec{p})$ around a momentum $\vec{p}_0$ which 
we do not specify for now: 
\be
E_j(\vec{p})\simeq E_j(\vec{p}_0)+\vec{v}_j(\vec{p}-\vec{p}_0)\,.
\label{eq:expan1}
\ee
An important point is that, strictly speaking, using such an expansion in the 
integral in (\ref{eq:Aj}) is only justified if $\vec{P}$ and $\vec{P}'$ are 
not too far from each other and if $\vec{p}_0={\cal O}(\vec{P},\vec{P}')$, 
which we will assume. As we shall see, under these conditions the final result 
will be insensitive to the choice of the expansion point $\vec{p}_0$. In 
eq.~(\ref{eq:expan1}) we have neglected higher order terms in the expansion; 
this corresponds to neglecting the spreading of the neutrino wave packets, 
which is unimportant for our argument. 
 
The Gaussian wave packets for the produced neutrino states are given by 
eq.~(\ref{eq:Gauss1}), and the detected wave packets have a similar form, 
with $\sigma_{pP}$ and $\vec{P}$ replaced by $\sigma_{pD}$ and $\vec{P}'$, 
respectively. Substituting these expressions and expansion (\ref{eq:expan1}) 
into eq.~(\ref{eq:Aj}) yields, after a simple Gaussian integration, 
\be
{\cal A}_j(T,\vec{L})\simeq 
\left(\frac{2\sigma_{pP}\sigma_{pD}}{\sigma_{pP}^2+\sigma_{pD}^2}\right)^{3/2} 
e^{-\frac{(\vec{P}-\vec{P}')^2}{4(\sigma_{pP}^2+\sigma_{pD}^2)}} 
e^{-iE_j(\vec{\bar{p}}) T + i \vec{\bar{p}} \vec{L}}
\,e^{-\frac{(\vec{L}-\vec{v}_j T)^2}{4\sigma_x^2}}\,.
\label{eq:Gauss2}
\ee 
Here 
\be
\bar{\vec{p}}\equiv \frac{\vec{P} \sigma_{pD}^2+\vec{P}' \sigma_{pP}^2}
{\sigma_{pP}^2+\sigma_{pD}^2}\,, \qquad\quad
\frac{1}{\sigma_p^2}\equiv\frac{1}{\sigma_{pP}^2}+
\frac{1}{\sigma_{pD}^2}\,,\qquad\quad
\sigma_x\equiv \frac{1}{2\sigma_p}\,.
\label{eq:not4}
\ee
and we have used the relation $E_j(\vec{p}_0)+\vec{v}_j(\bar{\vec{p}}-
\vec{p}_0)\simeq E_j(\bar{\vec{p}})$ which follows from (\ref{eq:expan1}). 
As can be seen from eq.~(\ref{eq:Gauss2}), the dependence on the 
expansion point $\vec{p}_0$ has completely disappeared from the 
amplitude. 

The meaning of the obtained result is quite transparent: as usual, the 
amplitude ${\cal A}_j(T, \vec{L})$ contains a normalization factor, a plane 
wave factor calculated at some mean momentum (in this case $\bar{\vec{p}}$) 
and the envelope factor 
$\exp[-(\vec{L}-\vec{v}_j T)^2/4\sigma_x^2]$. 
However, on top of this, it contains the extra factor $\exp[-(\vec{P}-
\vec{P}')^2/4(\sigma_{pP}^2+\sigma_{pD}^2)]$, which is a reflection of the 
approximate conservation of the mean neutrino momentum. It suppresses the 
amplitude unless the difference of the mean momenta of the produced and 
detected neutrino states is small compared to the effective total momentum 
width $(\sigma_{pP}^2+\sigma_{pD}^2)^{1/2}$. Note that this width is dominated 
by the larger of $\sigma_{pP}$ and $\sigma_{pD}$, unlike the 
momentum width $\sigma_p$ defined in (\ref{eq:not4}), which is dominated by 
the smaller of them. 

\vspace*{1.2mm} 
\noindent
{\em 2. Non-central collisions}

In our discussion of the external wave packet formalism, as well as in all 
other treatments of this topic we are aware of, it was assumed that the 
peaks of the wave packets of all the external particles participating in 
neutrino production (or detection) meet at the same space-time point.
In other words, it was assumed that at production these peaks are all located 
at the point $\vec{x}=\vec{x}_P$ at the same time $t=t_P$, 
whereas at detection the peaks of the wave packets of the external 
particles are all located at $\vec{x}=\vec{x}_D$ at the same time $t=t_D$. 
Is this always true and how crucial is this assumption?

While in some situations, such as neutrino production in decays of unstable 
particles, such an assumption may indeed be justified, this is not in general 
so when neutrinos are born in scattering processes, where the collisions 
of the wave packets in the initial state may be non-central. Let us 
consider non-central collisions assuming, as before, only two external 
particles at production and two at neutrino detection (the generalization to 
an arbitrary number of external particles is straightforward).  
One obvious consequence of this is
that the neutrino production and detection are only possible if the minimum 
distances between the peaks of the participating wave packets do not 
exceed significantly the sizes of, correspondingly, the production and 
detection regions.
Consider the production process. We shall now assume that the wave 
packets of the external particles $P_i$ and $P_f$ 
are given by expressions of the type (\ref{eq:psiA3}), with the phase
factors in the integrand being $e^{-iq(x-x_a)}$ and 
$e^{-i k(x-x_b)}$, respectively. This means that the peak of the wave 
packet describing $P_i$ is located at $\vec{x}=\vec{x}_a$ at the time 
$t=t_a$, and the peak of the wave packet of  $P_f$ is located at $\vec{x}=
\vec{x}_b$ at the time $t=t_b$. We will consider the positions of these peaks 
at the same time, i.e.\ we will take $t_a=t_b$. It is natural to choose $t_P$ 
to be the value of this common time corresponding to the minimum
distance between the two peaks, i.e.\ $t_a=t_b=t_P$, $|\vec{x}_a-\vec{x}_b|=
\min\{|\vec{x}_a(t)-\vec{x}_b(t)|\}$.%
\footnote{This is easily generalized to the case of more than two external 
particles. For instance, for three particles with peaks at $\vec{x}_a$, 
$\vec{x}_b$ and $\vec{x}_c$ the time $t_P$ can be chosen to correspond to the 
minimum of $(\vec{x}_a-\vec{x}_b)^2+(\vec{x}_a-\vec{x}_c)^2+
(\vec{x}_b-\vec{x}_c)^2$.}
The coordinate $\vec{x}_P$ can then be chosen to lie anywhere between 
$\vec{x}_a$ and $\vec{x}_b$ on the line connecting them; in particular, one 
can choose $\vec{x}_P=\vec{x}_a$ or $\vec{x}_P=\vec{x}_b$. 
Neutrino detection can be considered quite similarly. 

One can now repeat the calculations presented in Sec.~\ref{sec:QFT}, arriving 
at the transition amplitude that can again be written in the 
form~(\ref{eq:amp6}). However, the expressions for $\Phi_{jP}(p^0,\vec{p})$ 
and $\Phi_{jD}(p^0,\vec{p})$ in eq.~(\ref{eq:psi}) have to be modified: their 
integrands should be multiplied, respectively, by $e^{i \vec{q}(\vec{x}_P-
\vec{x}_a)-i \vec{k}(\vec{x}_P-\vec{x}_b)}$ and $e^{i \vec{q}'(\vec{x}_P-
\vec{x}_c)-i \vec{k}'(\vec{x}_P-\vec{x}_d)}$, where $\vec{x}_c$ and 
$\vec{x}_d$ are the positions of the peaks of the wave packets representing 
$D_i$ and $D_f$ at detection at the time $t_D$ when the distance between 
these two peaks reaches its minimum. This can be taken into account by 
redefining the momentum distribution functions of the wave packets according 
to 
\begin{align}
&f_{Pi}(\vec{q},\vec{Q})~\to~ f_{Pi}(\vec{q},\vec{Q}) e^{i \vec{q}
(\vec{x}_P-\vec{x}_a)}\,, \qquad \quad~
f_{Pf}(\vec{k},\vec{K}) ~\to~ f_{Pf}(\vec{k},\vec{K}) e^{i \vec{k}(\vec{x}_P
-\vec{x}_b)}\,, \nonumber \\
&f_{Di}(\vec{q}',\vec{Q}') ~\to~ f_{Di}(\vec{q}',\vec{Q}') e^{i \vec{q}'
(\vec{x}_P-\vec{x}_c)}\,,\qquad 
f_{Df}(\vec{k}',\vec{K}') ~\to~ f_{Df}(\vec{k}',\vec{K}') e^{i \vec{k}'
(\vec{x}_P-\vec{x}_d)}\,.
\label{eq:redef}
\end{align}
The newly defined momentum distribution functions share a crucial feature with
the old ones: they decrease rapidly when the deviation of the corresponding
momenta from their peak values exceed the relevant momentum uncertainties
(i.e.\ the widths of the peaks) $\sigma_{pPi}$, $\sigma_{pPf}$, $\sigma_{pDi}$
or $\sigma_{pDf}$. In addition, the new momentum distributions do not exhibit
fast oscillations when the momentum variations are smaller than, or of the
order of, the corresponding widths of the peaks. Indeed, consider neutrino
production. Our choice of $\vec{x}_P$ implies that $|\vec{x}_P-\vec{x}_a|\le
|\vec{x}_a-\vec{x}_b|$ and $|\vec{x}_P-\vec{x}_b|\le |\vec{x}_a-\vec{x}_b|$.
On the other hand, as we mentioned above, the production is only possible when
the distance between the peaks of the wave packets of the external particles
$|\vec{x}_a- \vec{x}_b|$ is smaller than, or of the order of, the size of the
production region, which is 
of the order of $[\max\{\sigma_{pPi},\,\sigma_{pPf}\}]^{-1}$.
A similar argument applies to neutrino detection. Therefore the variation of 
the exponents of the exponential factors in eq.~(\ref{eq:redef}) is $\lesssim 
1$ and these factors do not undergo fast oscillations across the peaks of the 
momentum distributions. 

From the above considerations it follows that the properties of the 
redefined momentum distributions are essentially the same as those of the 
old ones. All the results of the present paper therefore apply to the 
case of non-central collisions as well, if one substitutes the original 
momentum distribution functions by those redefined according to 
eq.~(\ref{eq:redef}).

%=======================================================================
\section{Discussion and summary\label{sec:disc}}
%=======================================================================

In this paper we have compared the quantum mechanical approach to neutrino 
oscillations, where neutrinos are described by wave packets, with the quantum 
field theoretical method, where they are represented by propagators connecting 
the neutrino production and detection vertices in a Feynman diagram, whereas 
the external particles are described by wave packets in order to localize the 
process in space and time. We have shown how the neutrino wave packets 
underlying the QM approach can be derived in QFT by comparing the QM and 
QFT expressions for the transition amplitude. 
Equivalently, the wave packet representing the emitted neutrino can 
be obtained as the convolution of the neutrino source (the production 
amplitude) with the retarded neutrino propagator, in accord with the well 
known result of QFT. Quite analogously, the wave packet of the detected  
neutrino can be obtained as the convolution of the neutrino detection 
amplitude with the advanced neutrino propagator, with the result 
then taken at the time corresponding to neutrino detection.  

We have studied the general properties of QFT-derived wave packets 
representing the produced neutrino states and demonstrated that the wave 
packets of mass eigenstates $\nu_j$ depend on the index $j$ only through the 
neutrino energy $E_j$, and that in the momentum representation they are 
given by the production amplitude multiplied by ``smeared delta functions'' 
describing approximate conservation of mean energy and mean momentum at 
production. The widths of these ``smeared delta functions'' are determined by 
the largest among the corresponding widths of the external particles 
involved in the neutrino production process. Similar conclusions apply to the wave 
packets of the detected neutrino states. 
 
We also identified the conditions under which general neutrino wave packets can 
be approximated by Gaussian ones. Using Gaussian wave packets as an example, 
we then demonstrated that the neutrino wave packets derived in QFT can be cast
into the form they are usually assumed to have in the QM 
formalism, provided that

(i) The momentum uncertainty of the QM approach is replaced by the effective 
one, which depends not only on the true momentum uncertainty at production 
(or detection), but also on the corresponding energy uncertainty, as well as 
on the neutrino velocity and the effective velocity of the neutrino production 
(or detection) region. Moreover, these momentum uncertainties are different 
in different directions, i.e.\ are anisotropic. 
The longitudinal effective momentum uncertainties of the produced and detected 
neutrino states are dominated by the energy uncertainties characterizing, 
respectively, the neutrino production and detection processes, whereas the 
transverse effective momentum uncertainties coincide with the 
corresponding true momentum uncertainties. 

(ii) The mean momentum of the neutrino state is shifted from its naively 
expected value. 

(iii) The wave packets of different mass eigenstates acquire (in general
different) extra overall factors.

\noindent
Thus, the simplistic QM wave packet approach may need QFT-motivated 
modifications; however, once these modifications have been done, one can still 
work within the QM framework without losing any essential physical content.   

We have also studied the energy uncertainties characterizing the neutrino 
wave packets in the case of unstable neutrino sources and have shown
that in general these uncertainties depend both on the decay rate of 
the parent particle and on the inverse time scale of the overlap of 
the wave packets of the external particles. The neutrino energy 
uncertainty is dominated by the larger of the two. 

In the last part of the paper, we have discussed in detail the normalization of 
the QM and QFT expressions for the oscillation probability 
$P_{\alpha\beta}(L)$. 
We have seen that in the QM framework $P_{\alpha\beta}(L)$ has to be 
normalized by hand in order to fulfill the unitarity relation $\sum_\beta 
P_{\alpha\beta}(L) = 1$. There are two reasons why this ad hoc procedure 
is unavoidable in the QM approach. First, as we have demonstrated, no 
independent normalization of the produced and detected neutrino states can 
lead to the correct normalization of the oscillation probability 
because the overlap integral of these states is always smaller than unity in 
the realistic case when their momentum distributions are different.  
The overlap integral depends on the 
characteristics of the produced and detected states in a non-factorizable way, 
and so the problem cannot be cured by just modifying the normalization 
of these states. Second, the QM formalism involves an integration over 
the unobserved difference of the neutrino detection and production times 
$T=t_D-t_P$, which leads to yet another undefined factor -- the time interval 
by which one has to divide the result in order to recover the correct 
dimension of the oscillation probability. In the QM method both these problems 
are solved by imposing unitarity of the oscillation probability by hand.   

We have demonstrated how the QFT approach avoids all the normalization problems
of the QM formalism and we have derived the conditions under which it naturally
leads to the correctly normalized oscillation probability that automatically
satisfies the unitarity condition. The conditions are that (1) neutrinos are
ultra-relativistic or quasi-degenerate in mass, and that, in addition, (2) the
differences $|p_j - p_k|$ between the momenta of different neutrino mass
eigenstates at fixed energy are much smaller than the widths of the neutrino
momentum distributions determined by the production and detection processes. If
these requirements are not fulfilled, the interaction rate cannot be factorized
into the production rate, propagation (oscillation) probability and detection
cross section, so that the oscillation probability is undefined. In that case
one would have to deal instead with the overall rate of the neutrino
production-propagation-detection process. 

The QFT approach also allows one to understand the physical meaning of 
the QM normalization recipe: By imposing unitarity by hand one implicitly  
rids the calculated transition probability of the probabilities of neutrino 
production and detection, thus extracting the sought oscillation probability. 
   
A comment on the the integration over $T$ is in order. 
Such an integration is involved in both the QM and QFT approaches to neutrino 
oscillations. 
In the QM framework, it has to be introduced to account for the fact that the
neutrino's time of flight is not measured (or at least not measured accurately
enough) in realistic experiments. At the same time, in our QFT treatment of
neutrino oscillations, it emerges naturally from the observation that in real
situations
one has to deal with continuous fluxes of incoming particles (or with 
ensembles of neutrino emitters and absorbers in the case of bound 
stationary initial states) rather than with individual acts of 
neutrino production, propagation and detection, in which single wave packets 
of the external particles of each type are involved. Can one still sensibly 
define an unintegrated oscillation probability $P_{\alpha\beta}(T, L)$ for 
such a single act?

We will argue now that the probability $P_{\alpha\beta}(T, L )$ is not a 
useful quantity since it is unmeasurable (or almost unmeasurable). An 
important point here is that in practice both $L$ and $T$ can only be 
measured with some accuracy.
If we consider the $T$-integrated probability $P_{\alpha\beta}(L)$ which 
depends (in addition to neutrino energy) only on $L$, then this quantity is 
well defined only if the error $\Delta L$ in the measurement of $L$ is small
compared to the variations of $L$ over which the probability changes 
significantly. This means that this error must be small compared to the 
neutrino oscillation length $l_{osc}$, and this condition is normally easily 
satisfied. If one considers the case when the time $T$ is measured, whereas 
the distance $L$ is not (even though it is hard to imagine how such a 
situation could be realized in practice), then one would have to integrate 
$P_{\alpha\beta}^{\rm tot}(T,L)$ over $L$, and the resulting probability 
would be a function of $T$. Then the situation would be similar -- the error 
$\Delta T$ in the determination of $T$ would have to be small compared to the 
change of $T$ over which $P_{\alpha\beta}(T)$ varies significantly, which is 
$\sim (l_{osc}/v)$ with $v$ the neutrino velocity. 

The situation would be quite different if one considers the unintegrated
probability $P_{\alpha\beta}(T, L )$ which depends on both $T$ and $L$ 
-- in this case the requirements on 
$\Delta L$ and $\Delta T$ would be by far more demanding. Indeed, 
$P_{\alpha\beta}(T, L )$ is substantially different from zero only when 
$|L-vT|\lesssim \sigma_x$, where 
$\sigma_x$ is the spatial length  of the neutrino wave packet. This means 
that for uncorrelated variations of $L$ and $T$ this probability 
varies significantly when $L$ 
changes by $\sigma_x$ (which can be extremely small) and $T$ changes by 
$(\sigma_x/v)$. The latter quantity is essentially given by the largest 
between the time scales of the neutrino production and detection 
processes. Thus, the un-integrated probability $P_{\alpha\beta}(T, L )$ 
can only be accurately measured if the distance $L$ is measured with an
accuracy better than the length of the neutrino wave packet and 
simultaneously the time between neutrino emission and detection is 
measured with an error that is small compared to the duration of these 
processes. Such a possibility appears rather unrealistic at the very 
least.  

It follows from our results that there are both intricate relations 
and important differences between the QM and QFT approaches to neutrino 
oscillations. 
In the following table, we compare the main features of these two 
approaches. 
\newpage
{\small
\newcounter{myitem}
\setcounter{myitem}{1}
\renewcommand{\arraystretch}{1.3}
\begin{longtable}{lp{0.45\textwidth}p{0.45\textwidth}}
  \toprule
    & \bf QM approach & \bf QFT approach \\
  \midrule

  \arabic{myitem}. \addtocounter{myitem}{1} &
  Simple and transparent description of neutrino oscillations.
  Neutrino production and detection processes are not properly 
  taken into account. Simplified description of neutrino energy and 
  momentum uncertainties. 
     &
  Most complete description of neutrino production, propagation and
  detection. 
  Accurate treatment of neutrino energy and 
  momentum uncertainties.   
  The formalism is more complicated than that of the QM approach. \\
  \arabic{myitem}. \addtocounter{myitem}{1} &
  Produced and detected neutrino states are flavour eigenstates, defined 
  according to
  \begin{align*}
    \ket{\nu_\alpha} = \sum_j U_{\alpha j}^* \ket{\nu_j}
  \end{align*}\vspace{-1cm}
     &
  Only mass eigenstates are considered. 
  (In fact, defining flavour eigenstates in QFT poses great difficulties 
  because they do not form a physically meaningful Fock space 
  \cite{Giunti:2003dg}.) \\
  \arabic{myitem}. \addtocounter{myitem}{1} &
  Mass eigenstates composing neutrino flavour eigenstates are described by 
  wave packets whose form is postulated rather than derived and the 
  parameters (momentum uncertainties) are estimated from the properties 
  of the production and detection processes. Derivation of neutrino wave 
  functions is not possible since they depend on the dynamics of neutrino 
  production and detection, and particle creation and annihilation 
  cannot be described in QM. 
    &
  Because neutrinos are only in the intermediate states, their wave functions 
  are not necessary for the formalism (but can be derived from the production 
  and detection amplitudes according to well defined rules). Wave functions 
  of the external particles accompanying neutrino production and detection have 
  to be known. Undetected external particles can be described by plane waves. \\
  \arabic{myitem}. \addtocounter{myitem}{1} &
  The oscillation amplitude is obtained by evolving the produced neutrino state
  in time and then projecting it onto the detected state.
     &
  The amplitude of the combined process of neutrino production, propagation,
  and detection is computed according to the Feynman rules. Time evolution of 
  the neutrino states in QM corresponds to their on-shell propagation in 
  QFT. Projection in QM corresponds to the integration over the 
  momentum of the intermediate neutrino in QFT. \\
  \arabic{myitem}. \addtocounter{myitem}{1} &
  The oscillation probability has to be normalized by hand by imposing the 
  unitarity condition. (The physical meaning
  and justification of this normalization procedure is elucidated in QFT.)
     &
  The oscillation probability $P_{\alpha\beta}(L)$ that is properly normalized 
  and satisfies the unitarity constraint is automatically obtained from the 
  formalism when neutrinos are ultra-relativistic or 
  quasi-degenerate and when, in addition, the momentum differences between
  different mass eigenstates at fixed energy are much smaller than the widths
  of the neutrino momentum distributions. Otherwise $P_{\alpha\beta}(L)$ is
  undefined.  \\

  \bottomrule
\end{longtable}
}

In summary, we have explicated the close relation between the quantum
mechanical and quantum field theoretical approaches to neutrino oscillations 
and have shown how QFT -- apart from providing expressions for oscillation
probabilities and event rates in its own right -- can be used to derive 
the input parameters required for the QM approach and to elucidate some QM 
procedures which were not properly justified or fully understood within that 
approach. 
We have also clarified several subtle points regarding neutrinos from unstable 
sources, the case of unequal mean momenta of the produced and detected 
neutrino states and the normalization of the oscillation probability.

\section*{Acknowledgements} The authors are grateful to Fedor Bezrukov and
Alexei Smirnov for very useful discussions. Fermilab is operated by Fermi
Research Alliance, LLC under Contract No.~DE-AC02-07CH11359 with the US
Department of Energy.

%=======================================================================
\appendix
\renewcommand{\theequation}{\thesection\arabic{equation}}
\appsection
\renewcommand{\thesection}{\Alph{section}}
\section*{Appendix \Alph{section}: 
Matching the QFT and QM wave packets in the 3-dimensional case}
%=======================================================================

We present here the full solutions of eqs.~(\ref{eq:shiftEq}) 
and~(\ref{eq:normtot}). 

The quantities $\alpha^{kl}$ and $\delta^k$ depend on the components of
two linearly independent vectors, $\vec{v}_j-\vec{v}_P$ and $\vec{v}_j$. 
We therefore seek the solution of eq.~(\ref{eq:shiftEq}) in the form 
\be
\delta^k=c_0 (v_j-v_P)^k + c_1 v_j^k\,.
\label{eq:deltaK}
\ee
Substituting this into eq.~(\ref{eq:shiftEq}) and comparing the 
coefficients of $(v_j-v_P)^l$ and of $v_j^l$ on both sides of the 
equality, one finds
\be
c_0=-\frac
{(E_j-E_P)[\lambda_P+\frac{E_j-E_P}{E_j}(1-\vec{v}_j^2)]}
{[\lambda_P+\frac{E_j-E_P}{E_j}(1-\vec{v}_j^2)]
[\lambda_P+(\vec{v}_j-\vec{v}_P)^2+\frac{E_j-E_P}{E_j}]
+\frac{E_j-E_P}{E_j}[(\vec{v}_j-\vec{v}_P) \vec{v}_j]^2}\,,
\label{eq:c0}
\ee
\be
c_1=c_0\,\frac{E_j-E_P}{E_j}\frac{[(\vec{v}_j-\vec{v}_P) \vec{v}_j]}
{[\lambda_P+\frac{E_j-E_P}{E_j}(1-\vec{v}_j^2)]}\,.
\label{eq:c1}
\ee
The parameter $\tilde{\gamma}_j$ can then be obtained from 
eq.~(\ref{eq:normtot}), which gives
\be
\tilde{\gamma}_j=\frac{(E_j-E_P)^2}{4\sigma_{eP}^2}+
\frac{(E_j-E_P)}{4\sigma_{eP}^2}\big[c_0(\vec{v}_j-\vec{v}_P)^2+c_1
[(\vec{v}_j-\vec{v}_P)\vec{v}_j]\big]\,. 
\label{eq:gamma1}
\ee
Substituting here the expressions for $c_0$ and $c_1$, one finds  
\be
\tilde{\gamma}_j=\frac{(E_j-E_P)^2}{4\sigma_{eP}^2}
\frac{
[\lambda_P+\frac{E_j-E_P}{E_j}(1-\vec{v}_j^2)] 
[\lambda_P+\frac{E_j-E_P}{E_j}]}
{[\lambda_P+\frac{E_j-E_P}{E_j}(1-\vec{v}_j^2)] 
[\lambda_P+(\vec{v}_j-\vec{v}_P)^2+\frac{E_j-E_P}{E_j}]
+\frac{E_j-E_P}{E_j}[(\vec{v}_j-\vec{v}_P)\vec{v}_j]^2}.
\label{eq:gamma2}
\ee

The above expressions simplify significantly if one makes use of the fact that 
$\frac{E_j-E_P}{E_j}$ is small (note that for relativistic neutrinos 
the above formulas contain one more small parameter, $(1-\vec{v}_j^2)$). 
In the leading order in $\frac{E_j-E_P}{E_j}$ the 
effective momentum uncertainties that are related to the eigenvalues of 
the matrix $\alpha^{kl}$ are given by eq.~(\ref{eq:sigmaPeff3})
of Sec.~(\ref{sec:wp-match}), whereas the shift of the mean momentum and 
the parameters $\tilde{\gamma}_j$ are given by eq.~(\ref{eq:deltaK2}). 

%=======================================================================
\appsection
\renewcommand{\thesection}{\Alph{section}}
\section*{Appendix \Alph{section}: 
Calculation of the integrals $I_1$, $I_2$ and $I_3$}
%=======================================================================

Here we calculate the integrals $I_1$, $I_2$ and $I_3$ that enter into 
eq.~(\ref{eq:probtot2}) in the limit of large $t$ ($t$ much larger than 
the time scales of the neutrino production and detection processes). 
For generality, we will not assume isotropy of neutrino emission and 
detection here. 
In the limit of large $t$ the integral $I_1$ is given by  
\begin{align}
I_1 = 
\frac{1}{64\pi^4 L^2} 
\sum_{j,k} U_{\alpha j}^* U_{\beta j}^{} U_{\alpha k}^{} U_{\beta k}^* 
\,\int_{-\infty}^{\infty}dT\,\int\!dE\, \int\!dE'\, 
\Phi_{P}(E,p_j\vec{l}) \Phi_{D}(E, p_j\vec{l})\,
\nonumber \\
\times \Phi^*_{P}(E', p_k'\vec{l}) \Phi^*_{D}(E', p_k'\vec{l})\,
\,(4E E')\,e^{-i(E-E')T+i (p_j-p_k') L}\,,
\label{eq:Int1a}
\end{align}
where we use the shorthand notation $p_k'\equiv p_k(E')$. 
This quantity coincides with the time-integrated probability 
$\tilde{P}_{\alpha\beta}^{\rm tot}(L)$ defined in eq.~(\ref{eq:Ptot1})
since $\int_{-\infty}^{\infty}dT\,e^{-i(E-E')T}=2\pi \delta(E-E')$.

Let us now show that in the limit of large $t$ the contributions of the 
integrals $I_2$ and $I_3$ to eq.~(\ref{eq:probtot2}) 
are negligible.   
We first notice that $I_3$ is essentially zero 
because so is $P_{\alpha\beta}^{\rm tot}(T, L)$ for $T<0$. 
From $I_3=0$ it follows that we can replace $I_2-I_3$ by $I_2+I_3$, 
which turns out to be easier to calculate. For this quantity 
eq.~(\ref{eq:probtot2}) yields $I_2+I_3= \sum_{j,k} U_{\alpha j}^* 
U_{\beta j}^{} U_{\alpha k}^{} U_{\beta k}^*\,B_{jk}$ with
\begin{align}
B_{jk}~\equiv~
\frac{1}{64\pi^4 L^2} \int_{-\infty}^{\infty}dT\,T\,\int\!dE\, \int\!dE'\, 
\Phi_{P}(E,p_j\vec{l}) \Phi_{D}(E, p_j\vec{l})\,
\nonumber \\
\times \Phi^*_{P}(E', p_k'\vec{l}) \Phi^*_{D}(E', p_k'\vec{l})\,
\,(4E E')\,e^{-i(E-E')T+i (p_j-p_k') L}\,. 
\label{eq:Bjk2}
\end{align}
Note that the integral here is similar to that in eq.~(\ref{eq:Int1a}), 
but, unlike the latter, contains $\int_{-\infty}^{\infty}dT\, T 
e^{-i(E-E')T} 
=-2\pi i \delta'(E-E')$, where the prime stands for differentiation with 
respect to $E'$. Because the integral over $E'$ of $f(E')\delta'(E-E')$ 
yields ${}-f'(E)$, one obtains 
\be
B_{jk}=
\frac{2\pi i}{64\pi^4 L^2} 
\int_{-\infty}^{\infty} dE\, C_j(E) \,\frac{d}{dE} C_k^*(E)\,,
\label{eq:Bjk3}
\ee
where $C_j(E)\equiv \Phi_{P}(E,p_j\vec{l}) \Phi_{D}(E, p_j\vec{l})\,(2E)
\,e^{-i E T+i p_j(E) L}$. 
Integrating in (\ref{eq:Bjk3}) by parts we find
\be
B_{jk}=
\frac{2\pi i}{64\pi^4 L^2} \big[
C_j(E) C_k^*(E)\big|_{-\infty}^\infty ~- \int_{-\infty}^{\infty} dE\, 
C_k^*(E)\,\frac{d}{dE} C_j(E)\big] \,.
\label{eq:Bjk4}
\ee
The first term on the right hand side vanishes because so do the functions 
$\Phi_{P,D}$ 
at $E\to \pm\infty$. Therefore eq.~(\ref{eq:Bjk4}) means $B_{jk}=-B_{kj}^*$. 
On the other hand, from the definition (\ref{eq:Bjk2}) of $B_{jk}$ it follows 
that $B_{jk}=B_{kj}^*$. Hence, $B_{jk}=0$ and $I_2+I_3$ vanishes.

\end{document}